\documentclass[preprint]{ptephy_v1}

\preprintnumber{YITP-23-131} 
\usepackage{hyperref}

\usepackage{hyperref}     
\usepackage{amsmath}       
\usepackage{amssymb} 
\usepackage{graphicx}
\usepackage{url}
\usepackage{enumerate}
\usepackage{color}
\usepackage{ulem}
 
\usepackage{float,wrapfig,color}

\def\eq#1{(\ref{#1})}
\def\s[#1\s]{\begin{align}\begin{split}#1\end{split}\end{align}}
\def\[#1\]{\begin{align}#1\end{align}}

\def\bpsip{{\bar \psi_\parallel}}
\def\bpsit{{\bar \psi_\perp}}
\def\bpsione{{\bar \psi_{\perp_1}}}
\def\bpsitwo{{\bar \psi_{\perp_2}}}
\def\psip{{\psi_\parallel}}
\def\psione{{\psi_{\perp_1}}}
\def\psitwo{{\psi_{\perp_2}}}

\def\bvphip{{\bar \varphi_\parallel}}
\def\bvphit{{\bar \varphi_\perp}}
\def\bvphione{{\bar \varphi_{\perp_1}}}
\def\bvphitwo{{\bar \varphi_{\perp_2}}}
\def\vphip{{\varphi_\parallel}}
\def\vphit{{\varphi_\perp}}
\def\vphione{{\varphi_{\perp_1}}}
\def\vphitwo{{\varphi_{\perp_2}}}

\def\phip{{\phi_\parallel}}
\def\phit{{\phi_\perp}}
\def\phione{{\phi_{\perp_1}}}
\def\phitwo{{\phi_{\perp_2}}}

\def\sigp{{\sigma_\parallel}}
\def\sigt{{\sigma_\perp}}
\def\sigone{{\sigma_{\perp_1}}}
\def\sigtwo{{\sigma_{\perp_2}}}

\def\tal{{\tilde\alpha}}
\def\tbe{{\tilde\beta}}
\def\nt{{n_\perp}}
\def\np{{n_\parallel}}
\def\tv{{\tilde v}}

\begin{document}

\title{Real eigenvector distributions of random tensors\\ 
with backgrounds and random deviations}


\author{Naoki Sasakura}
\affil{Yukawa Institute for Theoretical Physics, Kyoto University, \\
and \\
CGPQI, Yukawa Institute for Theoretical Physics, Kyoto University, \\
Kitashirakawa, Sakyo-ku, Kyoto 606-8502, Japan
 \email{sasakura@yukawa.kyoto-u.ac.jp}}

\begin{abstract}%
As in random matrix theories, eigenvector/value distributions 
are important quantities of random tensors in their applications. 
Recently, real eigenvector/value distributions of Gaussian random tensors 
have been explicitly computed by expressing them as partition functions of quantum field theories
with quartic interactions.
This procedure to compute distributions in random tensors is general, powerful and intuitive, 
because one can take advantage of well-developed techniques and knowledge of quantum field theories.
In this paper we extend the procedure to the cases that random tensors have mean backgrounds 
and eigenvector equations have random deviations. In particular,
we study in detail the case that the background is a rank-one tensor, 
namely, the case of a spiked tensor. 
We discuss the condition under which the background rank-one tensor has a visible peak in  
the eigenvector distribution. We obtain a threshold value, which agrees with a previous result
in the literature.
\end{abstract}

\subjectindex{A13, A45, B83, B86}

\maketitle

\section{Introduction}
\label{sec:introduction}
Eigenvalue distributions are important quantities in random matrix models.
The most well-known is the Wigner semi-circle law of the eigenvalue distribution, 
which models energy spectra of strongly interacting many-body systems \cite{Wigner}.  
Eigenvalue distributions are also used as 
an important technique in solving matrix models \cite{Brezin:1977sv,matrix}. Topological changes of eigenvalue
distributions provide insights into the QCD dynamics \cite{Gross:1980he,Wadia:1980cp}. 

It would be natural to ask how such knowledge about random matrices can be generalized to 
random tensors. 
Random tensor models \cite{Ambjorn:1990ge,Sasakura:1990fs,Godfrey:1990dt,Gurau:2009tw} 
were originally introduced to extend random matrix models, which are successful as two-dimensional
quantum gravity, to higher dimensional quantum gravity.
 Recently random tensor models also play interesting roles in various other subjects
 (See for instance \cite{Ouerfelli:2022rus}). 
While physically interesting matrices like hermitian 
can be one-to-one mapped to sets of eigenvalues by symmetry transformations,
this cannot be done in general for tensors.
However, we sometimes encounter what we may call tensor eigenvectors/values \cite{Qi,lim,cart,qibook}
in studies.
A well-known example is the distribution of the energy spectra of the spherical $p$-spin model 
\cite{pspin,pedestrians} for spin glasses, which was comprehensively analyzed in \cite{randommat}.
In fact this is the same problem as obtaining the real eigenvalue\footnote{More precisely, they are 
Z-eigenvalues in the terminology of \cite{Qi,qibook}.} distribution of a real symmetric random tensor.
Tensor eigenvector/value problems also appear in other contexts, such as 
AdS/CFT \cite{Biggs:2023mfn}, classical gravitational systems \cite{Evnin:2021buq}, and
applied mathematics for technologies \cite{qibook}.

Considering its wide appearance, it is worth effort to systematically understand
properties of tensor eigenvectors/values. Our focus is on their distributions for Gaussian random tensors.
Some interesting results have already been obtained
in the literature. In \cite{realnum1,realnum2} the expectation values of numbers of real eigenvalues of 
random tensors were computed. In \cite{Evnin:2020ddw} the maximum eigenvalues of random tensors 
were estimated in the large-$N$ limit\footnote{Throughout this paper, $N$ denotes the range of 
indices of tensors, namely, an index takes values, $1,2,\cdots,N$.}. 
In \cite{Gurau:2020ehg}, the Wigner semi-circle law was extended to a form for random tensors. 
In \cite{Sasakura:2022zwc,Sasakura:2022iqd,Sasakura:2022axo} 
the present author computed real eigenvalue distributions of random tensors by quantum field theoretical methods.

In the last works above by the present author, 
the procedure is to first rewrite the eigenvector problems to partition functions of quantum field theories 
with quartic interactions, and then compute the partition functions. 
There are some merits in this procedure;
it is general, powerful, and intuitive. As far as tensors have Gaussian distributions,
one can in principle extend the procedure to obtain quantum field theories of quartic interactions 
for a wide range of other tensor problems, such as complex eigenvalue/vector distributions, 
tensor rank decompositions, etc.  
Then, once such quantum field theories have been obtained, 
one can use various well-developed quantum field theoretical techniques,
such as Schwinger-Dyson equations as in \cite{Sasakura:2022iqd}, etc. 
Moreover, it is generally more intuitive to compute partition functions than 
to directly treat systems of eigenvector/value polynomial equations. For instance, in the large-$N$ analysis
of \cite{Sasakura:2022iqd}, there exists a phase transition point 
between perturbative and non-perturbative regimes of the quantum field theory, and 
this point corresponds to the edge of the eigenvalue distribution.

The purpose of the present paper is to apply this quantum field theoretical procedure to
a slightly different setup than the previous 
works \cite{Sasakura:2022zwc,Sasakura:2022iqd,Sasakura:2022axo}. 
We assume the random tensors have mean 
values, namely, backgrounds. 
This is a useful setup in the research of data analysis, in which backgrounds are 
signals and deviations around them are noises \cite{spike}. It is an important
question under what conditions signals can be recovered from data contaminated by 
noises \cite{spike,spiketrans,AMN}.  
We also introduce random deviations to eigenvector 
equations\footnote{This particular case will also be analyzed in detail in \cite{nicolas}.}. This simulates 
solving approximately eigenvector equations, for instance, by the Monte Carlo method or 
simulated annealing. 
As we will see, also in this generalized setup, the distributions can be rewritten as partition functions of 
quantum field theories with quartic interactions, and the partition functions can be computed explicitly, even
exactly in some cases. 

This paper is organized as follows. In Section~\ref{sec:tensoreigen}, we introduce a real eigenvector
equation with a tensor mean background and deviations to the equation, and obtain an integral expression 
of the eigenvector distribution. 
In Section~\ref{sec:signed}, we derive the quantum field theory expressing a ``signed" distribution of the eigenvectors. 
This distribution is not authentic but is weighted with an extra sign associated to each eigenvector. 
This distribution is easier to compute, because the quantum field theory contains only a pair of fermions.  
In particular, when the background is taken to be a rank-one tensor (a spiked tensor),
we obtain an exact expression of the distribution in terms of hypergeometric functions.
In Section~\ref{sec:dist} we derive the quantum field theory expression of the (authentic) distribution of the eigenvectors. 
In particular we explicitly derive the distribution for the spiked tensor case
 by using an approximation taking advantage of the quantum field theoretical expression.  
In Section~\ref{sec:comp}, we compare the expressions of the distributions obtained in the previous sections with
 Monte Carlo simulations. We obtain very good agreement, including for the case treated by the approximation.
In Section~\ref{sec:largen}, we consider the large-$N$ limit, especially paying attention to whether the 
rank-one tensor background has a visible peak in the distributions. 
We derive the scaling and the range of parameters in which
this happens. The threshold value is shown to agree with that of \cite{AMN}.
The last section is devoted to summary and future prospects.

\section{Real tensor eigenvector equation with backgrounds and deviations}
\label{sec:tensoreigen}
In this paper we restrict ourselves to order-three 
tensors\footnote{Namely, tensors have three indices.} for simplicity. 
We consider the following eigenvector equation \cite{Qi,lim,cart,qibook}
with a background tensor $Q$ and a deviation vector $\eta$,
\[
(Q_{abc}+C_{abc})  v_b v_c =v_a +\eta_a.
\label{eq:egeq}
\] 
Here the indices take $a,b,c=1,2,\ldots,N$, and 
repeated indices are assumed be summed over unless otherwise stated throughout this paper.  
We assume that $Q,C$ are real symmetric order-three tensors and $v,\eta$ are real vectors:
\s[
&Q_{abc}=Q_{bac}=Q_{bca} \in \mathbb{R}, \\ 
&C_{abc}=C_{bac}=C_{bca} \in \mathbb{R}, \\
&v_a , \ 
\eta_a \in \mathbb{R}.
\s]
While $Q$ is an externally given background tensor, $C_{abc}$ is a random tensor with Gaussian
distribution of a zero mean value. 
The vector $\eta$ describes a deviation of the eigenvector equation, and 
is a random real vector with Gaussian distribution of a zero mean value.
We will compute the distributions of $v$, namely the distributions of the real ``eigenvector" 
solutions to \eq{eq:egeq}. 
Note that, if we ignore the background $Q$ and the deviation $\eta$, 
the setup goes back to the cases previously studied in 
\cite{Sasakura:2022zwc,Sasakura:2022iqd,Sasakura:2022axo}.

For given $Q,C,\eta$, the distribution of $v$ is given by 
\s[
\rho(v,Q,C,\eta)&=\sum_{i=1}^{\# {\rm sol}(Q,C,\eta)} \prod_{a=1}^N \delta(v_a-v_a^i) \\
&= \left | \det M(v,Q,C) \right| \prod_{a=1}^N \delta \left(  v_a +\eta_a- (Q_{abc}+C_{abc})  v_b v_c  \right)
\label{eq:distdelta}
\s]
where $v^i\ (i=1,2,\ldots,\# {\rm sol}(Q,C,\eta))$ are all the real solutions to \eq{eq:egeq}, and 
$|\det M(v,Q,C)|$ is the absolute value of the determinant of the matrix, 
\[
M(v,Q,C)_{ab} =\frac{\partial}{\partial v_a} \left(v_b+\eta_b-(Q_{bcd}+C_{bcd})  v_c v_d\right)=\delta_{ab}-2 (Q_{abc}+C_{abc})  v_c,
\]
which is the Jacobian factor associated to the change of the variables of the delta functions
in \eq{eq:distdelta}.

When $C,\eta$ have Gaussian distributions with zero mean values, the eigenvector distributions are computed 
by taking the average over $C,\eta$:
 \s[
\rho(v,Q,\beta)&=\left\langle \rho(v,Q,C,\eta) \right\rangle_{C,\eta} \\
&=\frac{1}{AA'} 
\int_{\mathbb{R}^{\# C}} dC  \int_{\mathbb{R}^N} d\eta  \, e^{-\alpha C^2-\frac{1}{4 \beta} \eta^2} 
\left | \det M(v,Q,C) \right | \prod_{a=1}^N \delta \left(  v_a +\eta_a- (Q_{abc}+C_{abc})  v_b v_c  \right),
\label{eq:average}
\s]
where $\alpha,\beta>0$, $\# C$ is the number\footnote{Explicitly, $\# C=N(N+1)(N+2)/6$.}
of the independent components of $C$, $C^2=C_{abc}C_{abc}$, $\eta^2=\eta_a \eta_a$,
$A=\int_{\mathbb{R}^{\# C}} dC\, e^{-\alpha C^2}$, and  
$A'=\int_{\mathbb{R}^N} d\eta\, e^{-\frac{1}{4 \beta} \eta^2}$. 
Here a slightly complicated introduction of $\beta$ is for later convenience.
By using the well known formula, 
$\frac{1}{2 \pi}\int_\mathbb{R} d\lambda \, e^{i\lambda x}=\delta(x)$, 
the integration of the delta functions over $\eta$ 
in \eq{eq:average} can be rewritten as 
\[
\frac{1}{A'} \int_{\mathbb{R}^N} d\eta  \, e^{-\frac{1}{4 \beta} \eta^2} \prod_{a=1}^N 
\delta \left(  v_a +\eta_a- (Q_{abc}+C_{abc})  v_b v_c  \right)=\frac{1}{(2 \pi)^{N}} \int_{\mathbb{R}^N} d\lambda\,
e^{-\beta\lambda^2+i \lambda_a \left(v_a - (Q_{abc}+C_{abc})  v_b v_c\right)}.
\]
Therefore, by putting this into \eq{eq:average}, we obtain
\[
\rho(v,Q,\beta)=\frac{1}{(2 \pi)^N A} \int_{\mathbb{R}^{\# C}} dC \int_{\mathbb{R}^N} d\lambda \ 
\left | \det M(v,Q,C) \right |
e^{-\alpha C^2-\beta \lambda^2
+i \lambda_a ( v_a - (Q_{abc}+C_{abc})  v_b v_c )}.
\label{eq:rhovq}
\]

The part $\left | \det M(v,Q,C) \right |$ in \eq{eq:rhovq} needs a special care, because taking an absolute value 
is not an analytic function. In Section~\ref{sec:signed}, we will consider the case that 
we ignore taking the absolute value. This makes the problem easier and treatable by introducing 
only a pair of fermions, but is still non-trivial and interesting. 
In Section~\ref{sec:dist}, we will fully treat \eq{eq:rhovq} by introducing both bosons and fermions.   

\section{Signed distributions}
\label{sec:signed}
\subsection{Quantum field theory expression}
The quantity we will compute in this section is defined by ignoring taking the absolute value in \eq{eq:rhovq}:
\[
\rho^{\rm signed}(v,Q,\beta)=\frac{1}{(2 \pi)^N A} \int_{\mathbb{R}^{\# C}} dC \int_\mathbb{R^N} d\lambda 
\det \! M(v,Q,C) \, e^{-\alpha C^2-\beta \lambda^2
+i \lambda_a ( v_a - (Q_{abc}+C_{abc})  v_b v_c )}.
\label{eq:rhosigned}
\]
Following backward the derivation in Section~\ref{sec:tensoreigen}, the distribution corresponds to 
a ``signed" distribution, 
\s[
\rho^{\rm signed} (v,Q,C,\eta)&=\sum_{i=1}^{\# {\rm sol}(Q,C,\eta)}
{\rm sign}\left( \det \! M(v^i,Q,C) \right)  \prod_{a=1}^N \delta(v_a-v_a^i), 
\s]
which has an extra sign of $\det \! M(v^i,Q,C)$ dependent on each solution $v^i$, 
compared with \eq{eq:distdelta}.
Note that the quantity \eq{eq:rhosigned} is a generalization of the signed distribution computed 
in \cite{Sasakura:2022zwc} to the case with backgrounds and deviations. 
Though the quantity has no clear connections to \eq{eq:rhovq}, it provides a simpler playground, 
and we will obtain an exact final expression 
with the confluent hypergeometric functions of the second kind (or hermite polynomials). 

The determinant factor in \eq{eq:rhosigned} can easily be rewritten in a quantum field theoretical form by introducing 
a fermion pair, $\bar \psi_a, \psi_a\ (a=1,2,\cdots, N)$; 
$\det M=\int d\bar\psi d\psi \, e^{\bar\psi M \psi}$ \cite{zinn}. 
This technique to incorporate determinants in quantum field theories is common in 
treating disordered systems in statistical 
physics\footnote{See for instance \cite{fyodorov} and references therein.}.
Then \eq{eq:rhosigned} can be rewritten as 
\[
\rho^{\rm signed}(v,Q,\beta)=\frac{1}{(2 \pi)^N A} \int_{\mathbb{R}^{\# C}} dC \int_{\mathbb{R}^N} d\lambda 
\int d\bar \psi d\psi \, e^{S^{\rm signed}_{\rm bare}},
\label{eq:rhosigneds0}
\]
where
\[
S^{\rm signed}_{\rm bare}=-\alpha C^2-\beta \lambda^2
+i \lambda_a ( v_a - (Q_{abc}+C_{abc})  v_b v_c )+\bar \psi_ a 
\left( \delta_{ab}-2 (Q_{abc}+C_{abc})  v_c \right) \psi_b.
\label{eq:s0}
\]

Since $C$ and $\lambda$ appear at most quadratically in \eq{eq:s0},
they can be integrated out by Gaussian integrations.
We will first integrate over $C$ and then over $\lambda$. 
Though the integrations are straightforward, the actual computation is a little cumbersome,
because of the anti-commuting nature of the fermions and the necessity of symmetrization 
for the indices of $C_{abc}$. 
However, we can take a shortcut by taking some results from \cite{Sasakura:2022zwc}, 
where there are no $Q$ or $\eta$.
Now, new terms in $S^{\rm signed}_{\rm bare}$ compared to \cite{Sasakura:2022zwc}
are those depending on $Q$ and $\beta$, and are explicitly given by
\[
S_{\rm new}^{\rm signed}=
-\beta \lambda^2-i \lambda_a Q_{abc} v_b v_c-2 Q_{abc}\bar \psi_a \psi_b v_c.
\label{eq:snew}
\]
Since the new terms do not contain $C$, the integration over $C$ proceeds in the same way as in  \cite{Sasakura:2022zwc}.
This integration cancels the overall factor $A^{-1}$ in \eq{eq:rhosigneds0}, and also
generates various terms being added to the action.  
Collecting the terms depending on $\lambda$ among the generated ones, 
$i \lambda_a v_a$ in \eq{eq:s0}, and the terms depending on 
$\lambda$ in \eq{eq:snew}, we obtain the $\lambda$-dependent part of the action as
\[
S_{\lambda}^{\rm signed}=-\frac{v^4}{12 \alpha} B_{ab} \lambda_a \lambda_b
+i \lambda_a (v_a+D^{\rm signed}_a-D^Q_a),
\label{eq:slamsg}
\]
where $D^Q_a=Q_{abc}v_b v_c$, and 
$D^{\rm signed}$ can be taken from \cite{Sasakura:2022zwc}\footnote{$v_a+D_a^{\rm signed}$ corresponds to
$D_a$ of  \cite{Sasakura:2022zwc}.},
\[
D^{\rm signed}_a& 
=\frac{1}{3 \alpha} \left( \bar \psi_a\, \psi\cdot v\, v^2+\bar\psi\cdot v \,\psi_a\, v^2+\bar \psi \cdot v\, \psi \cdot v \, v_a\right).
\label{eq:defofdsgn}
\] 
Here we very frequently use an abusive notation $v^p:=|v|^p$ for simplicity throughout this paper,
since whether $v$ means vector or scalar quantities are always obvious from contexts. 
The matrix $B$ is given by 
\[
B=3 \left( 1+\frac{4 \alpha \beta}{v^4}\right) I_\parallel+ \left( 1+\frac{12 \alpha \beta}{v^4} \right) I_{\perp},
\label{eq:matb}
\]
where $I_{\parallel}$ and $I_\perp$ are the projection matrices 
to the parallel and the transverse subspaces against 
$v$: $I_\parallel{}_{ab}=v_a v_b/v^2, \ I_\perp{}_{ab}= \delta_{ab}-v_a v_b/v^2$.
Then the integration over $\lambda$ with the action \eq{eq:slamsg} generates an action,
\s[
\delta S^{\rm signed}_{\lambda}=&-\frac{N}{2} \log\frac{v^4}{12 \pi \alpha } -\frac{1}{2} \log \det B \\
&-\frac{3 \alpha}{v^4} \left(
(v_a+D^{\rm signed}_a)B^{-1}_{ab} (v_b+D_b^{\rm signed}) -2 (v_a+D_a^{\rm signed})B_{ab}^{-1} D^Q_b +D^Q_a B^{-1}_{ab} D^Q_b\right) , 
\label{eq:delslam}
\s]
where the inverse of $B$ is given by
\[
B^{-1}=\frac{b_\parallel}{3} I_\parallel +b_\perp I_\perp
\label{eq:binv}
\]
with
\s[
&b_\parallel=\frac{v^4}{v^4+4 \alpha \beta},\\
&b_\perp= \frac{v^4}{v^4+12 \alpha \beta}.
\label{eq:bs}
\s]

When we consider the case with $Q=\beta=0$, the distribution \eq{eq:rhosigneds0}
should agree with the previous result of \cite{Sasakura:2022zwc}.
Therefore it is enough for us to compute the additional part which appears only when $Q\neq 0$ or $\beta\neq 0$. 
By subtracting $\delta S^{\rm signed}_\lambda$ for $Q=\beta=0$ in \eq{eq:delslam} 
and using \eq{eq:binv}, we obtain
\s[
&\delta S^{\rm signed}_{\lambda}-\delta S^{\rm signed}_{\lambda}(Q=\beta=0)=
\frac{1}{2} \log b_\parallel +\frac{N-1}{2} \log b_\perp -\frac{3\alpha}{v^4}
\bigg[
\frac{b_\parallel-1}{3}(v+ D^{\rm signed}_\parallel)^2 \\
&
+(b_\perp-1) D^{\rm signed}_\perp \cdot D^{\rm signed}_\perp-\frac{2 b_\parallel}{3} (v+D^{\rm signed}_\parallel )D^Q_\parallel -2 b_\perp D^{\rm signed}_\perp \cdot D^Q_\perp 
+\frac{b_\parallel}{3} (D^Q_\parallel)^2+b_\perp D^Q_\perp \cdot D^Q_\perp \bigg],
\label{eq:subdels}
\s]
where $D^{\rm signed}_\parallel=v \cdot D^{\rm signed}/|v|, \ D^{\rm signed}_\perp=I_\perp D^{\rm signed},\ D^Q_\parallel=v \cdot D^Q/|v|, \ D^Q_\perp=I_\perp D^Q$.

The previous result in \cite{Sasakura:2022zwc} is given by  
\[
\rho^{\rm signed}(v,Q=0,\beta=0)= 3^{\frac{N-1}{2}} \pi^{-\frac{N}{2}} \alpha^{\frac{N}{2}} \int d\bar \psi d\psi
\, e^{S_{\bar  \psi \psi}},
\label{eq:rhosignold}
\]
where 
\[
S_{\bar  \psi \psi}=-\frac{\alpha}{v^2} -2N \log v +\bpsit \cdot \psi_\perp -\bpsip \psip-
\frac{v^2}{6\alpha} \left( \bpsit \cdot \psi_\perp \right)^2
\label{eq:sbarqbzero}
\]
with $\psip=v \cdot \psi/|v|,\  \psi_\perp=I_\perp \psi$, etc.
Adding \eq{eq:subdels} and the last term in \eq{eq:snew} to \eq{eq:sbarqbzero} and 
doing some straightforward computations, we finally obtain
\s[
\rho^{\rm signed} (v,Q,\beta)=&
3^\frac{N-1}{2} \pi^{-\frac{N}{2}} \alpha^{\frac{N}{2}} (v^4+4 \alpha \beta)^{-\frac{1}{2}} 
(v^4+12 \alpha \beta)^{-\frac{N-1}{2}} \exp\left[- \frac{\alpha v^2}{v^4+4 \alpha \beta} \right]
\\
&
\cdot \exp\left[
\frac{2 \alpha b_\parallel v D^Q_\parallel-\alpha b_\parallel (D^Q_\parallel)^2 
-3 \alpha b_\perp D^Q_\perp\cdot D^Q_\perp }{v^4} 
 \right] 
\int d\bar \psi d\psi\, e^{S^{\rm signed}}, 
\label{eq:rhosignfinal}
\s]
where
\s[
S^{\rm signed}&=
 \left(-2 b_\parallel+1+\frac{2 b_\parallel D^Q_\parallel}{v} \right) \bpsip \psip 
 +\frac{2 b_\perp}{v}  D^Q_\perp \cdot \left( \bpsit \psip  +\bpsip \psi_\perp \right) +\bpsit \cdot \psi_\perp\\
 &\ \ 
 -2 Q_{abc}\bar \psi_a \psi_b v_c
+\frac{2 v^2(b_\perp-1)}{3 \alpha} \bpsip \psip \bpsit \cdot \psi_\perp 
-\frac{v^2}{6\alpha} \left( \bpsit \cdot \psi_\perp \right)^2.
\label{eq:ssignfinal}
\s]
Some details of the derivation are explained in Appendix~\ref{app:dels}.

\subsection{Rank-one $Q$}
\label{sec:rankonesign}
To study the formula \eq{eq:rhosignfinal} with \eq{eq:ssignfinal} 
more explicitly, let us consider the case that $Q$ is a rank-one tensor,
\[
Q_{abc}= q\, n_a n_b n_c,
\label{eq:rankone}
\]
where $q$ is real and $n$ is a normalized real vector ($|n|=1$).
This is a setup called a spiked tensor \cite{spike}.

In the general situation, the vector $n$ is a linear combination of $v$ and another vector $n_1$, which is a normalized vector transverse to $v$ (namely, $v\cdot n_1=0,\, |n_1|=1$). Then the transverse subspace to $v$ 
can further be divided into the subspace parallel to $n_1$ and the $N-2$-dimensional
subspace transverse to both $v$ and $n_1$.  
We denote the projector to the latter by $I_{\perp_2}$. Then the transverse fermions, $\bpsit, \psi_\perp$,
can further be decomposed into $\bpsione=n_1\cdot \bar \psi$ and $\bpsitwo=I_{\perp_{2}} \bar \psi$ and 
similarly for $\psi_\perp$. Note that $\bpsit\cdot \psi_\perp=\bpsione \psione+\bpsitwo\cdot \psitwo$, etc.

For \eq{eq:rankone}, $D^Q_\parallel=q v^2 n_\parallel^3,\ D^Q_\perp = q v^2 n_\parallel^2 n_\perp n_1$, where
$n_\parallel=v\cdot n /|v|,\ n_\perp=n_1\cdot n $. We also notice 
\s[
Q_{abc}v_c \bar \psi_a \psi_b&=q n_a n_b n_c v_a \psi_b \psi_c = q v n_\parallel^3 \bpsip \psip +q v n_\parallel^2 n_\perp(\bpsip \psione
+\bpsione \psip)+q v n_\parallel n_\perp^2 \bpsione \psione.
\s]
Putting these into \eq{eq:rhosignfinal} and \eq{eq:ssignfinal}, we obtain
\s[
\rho^{\rm signed}_{\rm spiked}(v,n,q,\beta)=&
3^\frac{N-1}{2} \pi^{-\frac{N}{2}} \alpha^{\frac{N}{2}} (v^4+4 \alpha \beta)^{-\frac{1}{2}} 
(v^4+12 \alpha \beta)^{-\frac{N-1}{2}} \\
&
\cdot \exp\left[ \frac{-\alpha v^2+2 \alpha q v^3 n_\parallel^3-\alpha q^2 v^4 n_\parallel^6}{v^4+4 \alpha \beta}
-\frac{3 \alpha q^2 v^4 n_\parallel^4 n_\perp^2 }{v^4+12 \alpha \beta}\right] 
\int d\bar \psi d\psi\, e^{S_{\rm spiked}^{\rm signed}},
\label{eq:rhopen}
\s]
where
\s[
S_{\rm spiked}^{\rm signed}&=
 -\left( \frac{v^4-4 \alpha \beta}{v^4+4 \alpha\beta} + \frac{8 \alpha \beta q v n_\parallel^3}{v^4+4 \alpha \beta}\right) \bpsip \psip
 -\frac{24 \alpha \beta q v n_\parallel^2 n_\perp}{v^4+12 \alpha\beta} 
\left(\bpsip \psione+\bpsione \psip \right) +(1-2 q v n_\parallel n_\perp^2) \bpsione \psione\\
&\ \ 
+
\bpsitwo \cdot \psitwo-\frac{8 \beta v^2}{v^4+12 \alpha \beta} \bpsip \psip \left( \bpsione \psione + \bpsitwo \cdot \psitwo \right)-\frac{v^2}{6\alpha} \left( \bpsione \psione + \bpsitwo \cdot \psitwo \right)^2. 
\label{eq:spen}
\s]

It is not difficult to explicitly compute the fermion integration in \eq{eq:rhopen}. 
As is shown in Appendix~\ref{app:delofU}, we obtain
\s[
\int d\bar \psi d\psi\, e^{S_{\rm spiked}^{\rm signed}}= 
2^{N-6} (-d_2)^{\frac{N-5}{2}} \bigg[&-8 d_2 (-b_2^2 + d_1 + b_3 (b_1 + d_1) + 
      2 b_1 d_2)\, U\left(\frac{3 - N}{2}, \frac{3}{2}, -\frac{1}{4 d_2}\right) \\
      &+ 2 (N-3) 
       (b_3 d_1 + 2 b_1 d_2 + 6 d_1 d_2)\, U\left(\frac{5 -N}{2}, 
       \frac{5}{2}, -\frac{1}{4 d_2}\right)  \\
       &- d_1 (N-3) (N-5)\, U\left(\frac{7- N}{2},\frac{7}{2}, -\frac{1}{4 d_2}\right)\bigg],
\label{eq:exact}       
\s]
where $U$ denotes the confluent hypergeometric function of the second kind, and 
$b_i,d_i$ are the coefficients of the terms in \eq{eq:spen}:
\s[
&b_1=-\left( \frac{v^4-4 \alpha \beta}{v^4+4 \alpha\beta} + \frac{8 \alpha \beta q v n_\parallel^3}{v^4+4 \alpha \beta}\right),\ 
b_2=-\frac{24 \alpha \beta q v n_\parallel^2 n_\perp}{v^4+12 \alpha\beta} , \ 
b_3=1-2 q v n_\parallel n_\perp^2,\\
&d_1=-\frac{8 \beta v^2}{v^4+12 \alpha \beta},\ 
d_2=-\frac{v^2}{6\alpha}.
\s]
The result \eq{eq:rhopen} with \eq{eq:exact} gives the exact expression of the signed distribution.

\section{Distributions}
\label{sec:dist}

\subsection{Quantum field theory expression}
In this subsection we compute the (authentic) distribution by considering 
the determinant factor $| \det M |$ as it is. 
We take the same procedure as 
was employed in \cite{Sasakura:2022axo}. We first introduce bosons and fermions to rewrite $| \det M |$:
\s[
|\det M | &=\lim_{\epsilon\rightarrow +0} \frac{\det (M^2 + \epsilon I)}{\sqrt{\det (M^2+\epsilon I )}}\\
&=(-\pi)^{-N} \int  d\bar \psi d\psi d\bar \varphi d\varphi d\phi d\sigma\, 
e^{-\sigma^2 -2 i \sigma M \phi-\epsilon \phi^2 
-\bar \varphi \varphi -\bar \psi M \varphi -\bar \varphi M \psi + \epsilon \bar \psi \psi},
\s]
where $I$ is an identity matrix of $N$-by-$N$, 
$\phi_a,\sigma_a$ are real bosons, $\bar \psi_a,\psi_a,\bar \varphi_a,\varphi_a$ are fermions, 
and $\bar \psi \psi=\bar \psi_a \psi_a$, etc. 
Here we have introduced a positive infinitesimal parameter $\epsilon$ to regularize the expression,
since $M$ may have zero eigenvalues.  
As in the second line, writing the limit is suppressed to simplify the notation hereafter, assuming
implicitly taking this limit at ends of computations. In fact the limit turns out to be straightforward
in all the computations of this paper. 
We have introduced two sets of bosons and fermions to make the exponent 
linear in $C$ ($M$ contains $C$ linearly) for later convenience of the integration over $C$. 
By performing similar processes as in Section~\ref{sec:signed}, we obtain 
\[
\rho(v,Q,\beta)=\frac{(-1)^N}{ 2^N \pi^{2N} A}
 \int dC d\lambda  d\bar \psi d\psi d\bar \varphi d\varphi d\phi d\sigma\, e^{S_{\rm bare}}, 
\]
where 
\s[
S_{\rm bare}=&-\alpha C^2-\beta \lambda^2 +i \lambda_a(v_a-(C_{abc}+Q_{abc})v_bv_c) \\
&-\sigma^2-2 i \sigma_a \left( \delta_{ab}-2 (Q_{abc}+C_{abc})  v_c \right) \phi_b-\epsilon \phi^2 \\
&-\bar\varphi\varphi-\bar \psi_a \left( \delta_{ab}-2 (Q_{abc}+C_{abc})  v_c \right) \varphi_b 
-\bar \varphi_a \left( \delta_{ab}-2 (Q_{abc}+C_{abc})  v_c \right) \psi_b+\epsilon \bar \psi \psi.
\label{eq:sbare}
\s]
As in Section~\ref{sec:signed}, there are no new terms depending on $C$ compared with 
the previous case for $Q=\beta=0$ in \cite{Sasakura:2022axo}, and therefore the integration over $C$ can be performed as in the previous computation there. 
Then we obtain a similar form of the action for $\lambda$ as in Section~\ref{sec:signed}:
\[
S_{\lambda}=-\frac{v^4}{12 \alpha} \lambda_a B_{ab} \lambda_b + i \lambda_a (v_a-D_a-D_a^Q),
\label{eq:slam}
\]
where $B,D^Q$ are already defined in \eq{eq:matb} and below \eq{eq:slamsg}, respectively.
Here $D$  can be taken from \cite{Sasakura:2022axo}\footnote{Here $D$ is the sum $D+\tilde D$ of 
\cite{Sasakura:2022axo}.}:
\s[
D_a&=\frac{v^3}{3 \alpha} \Big[ (\bpsip \vphip +\bvphip \psip)\hat v_a +
\bar \psi_a \vphip+ \bpsip \varphi_a+ \bar \varphi_a \psip+\bar
\vphip \psi_a+2  i \left(\hat v_a \sigp \phip+\sigma_a \phip+\sigp  \phi_a \right)\Big],
\label{eq:defofd}
\s]
where $\hat v_a=v_a/|v|$.
Comparing \eq{eq:slam} with \eq{eq:slamsg}, the change is to replace
$D^{\rm signed}$ with $-D$. By using \eq{eq:subdels} with this replacement
and adding the $Q$-dependent but $\lambda$-independent terms in \eq{eq:sbare},
we obtain
\s[
\rho(v,Q,\beta)=&
3^\frac{N-1}{2} \pi^{-\frac{3N}{2}} \alpha^{\frac{N}{2}} (v^4+4 \alpha \beta)^{-\frac{1}{2}} 
(v^4+12 \alpha \beta)^{-\frac{N-1}{2}} 
\exp\left[- \frac{\alpha v^2}{v^4+4 \alpha \beta} \right]
\\
&
\cdot \exp\left[
\frac{2 \alpha b_\parallel v D^Q_\parallel-\alpha b_\parallel (D^Q_\parallel)^2 -3 \alpha b_\perp D^Q_\perp\cdot
D^Q_\perp }{v^4} 
 \right] 
 Z,
\label{eq:rhodist}
\s]
where $Z$ is a partition function of a quantum field theory,
\[
Z=(-1)^N \int d\bar \psi \cdots d\sigma\, e^{S_0+S_{Q,\beta}}.
\label{eq:defofzdist}
\] 
Here $S_0$ is the former result in \cite{Sasakura:2022axo}
 corresponding to $Q=\beta=0$, which is explicitly given in Appendix~\ref{app:s0}, and
\s[
S_{Q,\beta}=&\frac{2 \alpha(b_\parallel-1) v -2\alpha b_\parallel D^Q_\parallel}{v^4} D_\parallel 
-\frac{6 \alpha b_\perp}{v^4} D_\perp\cdot D^Q_\perp 
+2 Q_{abc}v_c \left(\bar \psi_a \varphi_b +\bar \varphi_a \psi_b +2 i \sigma_a \phi_b\right) \\
&-\frac{\alpha(b_\parallel-1)}{v^4} D_\parallel^2-\frac{3 \alpha(b_\perp-1)}{v^4} D_\perp\cdot D_\perp,
\label{eq:sqbe}
\s]
where $D_\parallel=v\cdot D/|v|,\ D_\perp=I_\perp D$. Note that the first three terms are some corrections to the 
kinetic terms, and the latter to the four-interaction terms. As for $D_\parallel$ and $D_\perp$, 
we have more explicit expressions from \eq{eq:defofd},
\s[
&D_\parallel=\frac{v^3}{\alpha}\left( \bpsip \vphip+ \bvphip \psip+2 i \sigp \phip \right), \\
&D_\perp=\frac{v^3}{3\alpha} \left( 
\bpsit \vphip + \bpsip \vphit + \bvphit \psip+\bvphip \psi_\perp+2i (\sigp \phit+\sigt \phip)\right).
\s]

The four-interaction terms in \eq{eq:sqbe} have the form of self-products. 
One can make it quadratic by using 
the formula $\frac{1}{\sqrt{\pi}}\int_\mathbb{R} dg\, e^{-g^2 +2A g}= e^{A^2}$. The result is 
\[
Z=(-1)^N \pi^{-\frac{N}{2}}
\int dg_\parallel dg_\perp d\bar \psi \cdots d\sigma  \, e^{S_0+S_{Q,\beta,g}},
\label{eq:relwithg}
\] 
where $g_\parallel$ is one dimensional, $g_\perp$ is $N-1$ dimensional, and\footnote{Note that $b_\parallel,b_\perp<1$.} 
\s[
S_{Q,\beta,g}=&-g_\parallel^2-g_\perp^2+
\left( \frac{2 \alpha(b_\parallel-1) v -2\alpha b_\parallel D^Q_\parallel}{v^4} 
+\frac{2 \sqrt{\alpha(1-b_\parallel)}}{v^2} g_\parallel
\right)
D_\parallel 
-\frac{6 \alpha b_\perp}{v^4} D_\perp\cdot D^Q_\perp
\\
&+\frac{2 \sqrt{3 \alpha (1-b_\perp)}}{v^2} D_\perp \cdot g_\perp 
+2 Q_{abc}v_c \left(\bar \psi_a \varphi_b +\bar \varphi_a \psi_b +2 i \sigma_a \phi_b\right),
\label{eq:sqbeg}
\s]
which contains only quadratic terms of the fields.

\subsection{Rank-one $Q$}
\label{sec:rankone}
In this subsection we consider the rank-one tensor $Q$ in \eq{eq:rankone} to explicitly perform
the integration over the fields in \eq{eq:rhodist}. 

\subsubsection{A general formula}
By putting \eq{eq:rankone} into \eq{eq:rhodist}, one obtains
\s[
\rho(v,Q,\beta)=&
3^\frac{N-1}{2} \pi^{-\frac{3N}{2}} \alpha^{\frac{N}{2}} (v^4+4 \alpha \beta)^{-\frac{1}{2}} 
(v^4+12 \alpha \beta)^{-\frac{N-1}{2}} 
\exp\left[- \frac{\alpha v^2}{v^4+4 \alpha \beta} \right]
\\
&
\cdot \exp\left[
\frac{2 \alpha q v^3 n_\parallel^3-\alpha  q^2 v^4 n_\parallel^6}{v^4+4 \alpha \beta} -\frac{3 \alpha q^2 v^4 n_\parallel^4 n_\perp^2 }{v^4+12 \alpha \beta}
 \right] 
 Z,
 \label{eq:rhodistspiked}
\s]
where the partition function $Z$ can be 
computed either by \eq{eq:defofzdist} with \eq{eq:sqbe} or by \eq{eq:relwithg}
with \eq{eq:sqbeg}.

Let us first put \eq{eq:rankone} into \eq{eq:sqbe}. After a lengthy but straightforward computation using 
the same decomposition as in Section~\ref{sec:rankonesign},
we get
\s[
S_{q,n,\beta} := &S_{Q=qnnn,\beta} \\
=& 2 (q v n_\parallel^3-1) (1-b_\parallel)\left(
\bpsip \vphip+\bvphip \psip+2 i \sigp \phip
\right) \\
&+2 q v n_\parallel^2 n_\perp (1-b_\perp)
\left(\bpsione \vphip + \bpsip \vphione+ \bvphione \psip+\bvphip \psione+2i (\sigp \phione+\sigone \phip)\right) \\
&
+2 q v n_\parallel n_\perp^2 \left(
\bpsione \vphione+\bvphione \psione+2 i \sigone \phione
\right)\\
&+\frac{8\beta v^2}{v^4+4 \alpha \beta}\left(
-\bpsip \psip \bvphip \vphip+ 2 i (\bpsip \vphip
+\bvphip \psip) \sigp \phip -2 \sigp^2 \phip^2
\right)\\
&+\frac{8 \beta v^2}{v^4+12 \alpha \beta}
\bigg(
\bpsit \cdot \bvphit \psip \vphip +\psi_\perp\cdot \vphit \bpsip \bvphip -\bpsit \cdot \psi_\perp \bvphip \vphip
-\bpsit\cdot \vphit \bpsip \vphip \\
&\hspace{3cm}-\bvphit \cdot \vphit
\bpsip \psip -\bvphit\cdot \psi_\perp \bvphip \psip\\
&\hspace{3cm} +2 i \left(
\bpsit \vphip + \bpsip \vphit + \bvphit \psip+\bvphip \psi_\perp) \cdot (\sigp \phit+\sigt \phip\right) \\
&\hspace{3cm}-2 \left( 
\sigp^2 \phit\cdot \phit+\phip^2 \sigt\cdot \sigt
+2 \sigp \phip \phit \cdot \sigt
\right)
\bigg).
\label{eq:sqnbe}
\s]

As for \eq{eq:sqbeg}, we obtain
\s[
S_{q,n,\beta,g} := &S_{Q=qnnn,\beta,g} \\
=&-g_\parallel^2-g_\perp^2+
2 \left( (q v n_\parallel^3-1) (1-b_\parallel)
+  v \sqrt{\frac{1-b_\parallel}{\alpha}} g_\parallel\right) \left(
\bpsip \vphip+\bvphip \psip+2 i \sigp \phip
\right) \\
&+\left(2 q v n_\parallel^2 n_\perp (1-b_\perp)\right)
\left(\bpsione \vphip + \bpsip \vphione+ \bvphione \psip+\bvphip \psione+2i (\sigp \phione+\sigone \phip)\right) \\
&
+2 q v n_\parallel n_\perp^2 \left(
\bpsione \vphione+\bvphione \psione+2 i \sigone \phione
\right) \\
&
+2v \sqrt{\frac{1-b_\perp}{3\alpha}} g_\perp\cdot \left( 
\bpsit \vphip + \bpsip \vphit + \bvphit \psip+\bvphip \psi_\perp+2i (\sigp \phit+\sigt \phip)\right). 
\label{eq:swithg}
\s]

In the following subsections, we will consider $N=1$, $N=2$ and large-$N$ cases.

\subsubsection{$N=1$}
In this case we ignore all the transverse components, and also set $n_\parallel=1$.
By putting these to \eq{eq:rhodist}, \eq{eq:relwithg}, \eq{eq:swithg} and \eq{eq:s0form}, and doing some 
straightforward computations, we obtain
\[
\rho(v,q,\beta)=
\pi^{-1} \alpha^\frac{1}{2} (v^4+4 \alpha \beta)^{-\frac{1}{2}} 
\exp\left[ 
\frac{-\alpha v^2 +2\alpha q v^3 -\alpha q^2 v^4}{v^4+4 \alpha \beta} \right]
\left( \sqrt{\pi}a\, {\rm Erf}\left( \frac{a}{b} \right) +b\, e^{-\frac{a^2}{b^2}} \right),
\label{eq:distneq1}
\]
where
\s[
&a=1+2 (q v-1) (1-b_\parallel),\\
&b=2 v \sqrt{\frac{(1-b_\parallel)}{\alpha}}.
\label{eq:ab}
\s]
The details of the derivation are given in Appendix~\ref{app:neq1}.

\subsubsection{$N=2$}
\label{sec:neq2}
In this case the transverse direction is exhausted by one-dimension, namely, $\perp=\perp_1$
and $\perp_2$ is null. 
A special fact about this case is that the four-interaction terms in \eq{eq:s0form} have a form of a square:
\[
V_F+V_B+V_{BF}=\frac{v^2}{3 \alpha} \left( 
\bpsione \vphione+\bvphione \psione + 2 i \sigone \phione 
\right)^2.
\]
Therefore we can rewrite this part of the action as
\[
e^{V_F+V_B+V_{BF}}=\frac{1}{\sqrt{\pi}} \int dg \, e^{-g^2 + 2v g \left( 
\bpsione \vphione+\bvphione \psione + 2 i \sigone \phione
\right)/\sqrt{3\alpha}},
\]
whose exponent contains only quadratic terms of the fields. 
Using this for \eq{eq:relwithg}, \eq{eq:swithg} and \eq{eq:s0form},
we obtain
\[
Z_{N=2}=\pi^{-\frac{3}{2}} \int dg_1 dg_2 dg_3 
\int d\bar \psi\cdots d\sigma\, e^{-g_1^2-g_2^2-g_3^2+K_{\parallel \perp_1}},
\label{eq:zneq2}
\]
where 
\s[
K_{\parallel\perp_1}=&-\bvphip \vphip +\epsilon \bpsip \psip -\sigp^2 -\epsilon \phip^2-\bvphione \vphione +\epsilon \bpsione \psione -\sigone^2 -\epsilon \phione^2 \\
&+a_1 \left(\bpsip \vphip +\bvphip \psip+2 i \sigp \phip\right) \\
&+a_2 \left( \bpsip \vphione+\bpsione \vphip+\bvphip \psione+\bvphione \psip+2 i \left( \sigp \phione +\sigone \phip \right)\right) \\
&+a_3 \left(\bpsione \vphione +\bvphione \psione+2 i \sigone \phione\right)
\label{eq:kineticp1}
\s]
with
\s[
&a_1=2 b_\parallel-1+2q v  (1-b_\parallel) n_\parallel^3+2 v \sqrt{\frac{1-b_\parallel}{\alpha}} g_1,\\
&a_2=2 q v (1-b_\perp) n_\parallel^2 n_\perp+2 v \sqrt{\frac{1-b_\perp}{3 \alpha}} g_2,\\
&a_3=-1+2 qv n_\parallel n_\perp^2+2 v \sqrt{\frac{1}{3 \alpha}}g_3.
\label{eq:defofa}
\s]
Then the integration \eq{eq:zneq2} over the fields generates a square root of a determinant, 
and we obtain
\[
Z_{N=2}=\sqrt{\pi} \int dg_1 dg_2 dg_3\, e^{-g_1^2-g_2^2-g_3^2} \left| a_2^2-a_1a_3 \right|.
\]

\subsubsection{Large $N$}
\label{subsec:largen}
For $N>2$ we will not obtain exact expressions of the distributions. We will rather 
obtain an expression which is a good approximation for large $N$. 
For large $N$ the degrees of freedom carried by the $\perp_2$  fields 
will dominate over those of the $\parallel\perp_1$ fields, since the former is $(N-2)$-dimensional, 
while the latter is 2-dimensional.
Therefore the dynamics of the $\perp_2$ fields can well be determined by 
themselves with little effects from the $\parallel\perp_1$ fields, which may be ignored in the large-$N$ limit.
Then the dynamics of the $\parallel\perp_1$ fields may be computed in the backgrounds of  
the $\perp_2$ fields, which can well be approximated by their classical values because of their large number
of degrees of freedom for large $N$.

More precisely, our approximation is given by 
\[
Z=Z_{\perp_2} \, Z_{\parallel\perp_1}(R).
\label{eq:zapp}
\]
Here $Z_{\perp_2}$ is the partition function determined solely by the $\perp_2$ fields,
\[
Z_{\perp_2}=(-1)^{N-2} \int d\bpsitwo \cdots d\sigtwo \, e^{S_{\perp_2}},
\label{eq:zapp2}
\] 
where $S_{\perp_2}$ 
is the collection of the terms which contain only the $\perp_2$ fields in 
\eq{eq:s0v} with \eq{eq:s0form}\footnote{For instance, 
we include $\bar\psi_{\perp_2} \cdot \psi_{\perp_2}\bar\varphi_{\perp_2} \cdot \varphi_{\perp_2}$ but
ignore $\bar\psi_{\perp_2}\cdot \psi_{\perp_2}\bar\varphi_{\perp_1} \varphi_{\perp_1}$,
$\bar\psi_{\perp_2}\cdot \psi_{\perp_2}\bar\varphi_{\parallel} \varphi_{\parallel}$, etc.,
because of the reason mentioned in the first paragraph. The ignored terms will be considered in $Z_{\parallel \perp_1}$.}.
The computation of the partition function $Z_{\perp_2}$ is the same as that in the previous paper 
\cite{Sasakura:2022axo}, because $S_{\perp_2}$ has the same form as the action of the transverse directions
there\footnote{But note the difference of the dimensions of $\perp_2$ here and $\perp$ 
in \cite{Sasakura:2022axo}, where the former is $N-2$, while the latter is $N-1$. 
Therefore when we take a result from \cite{Sasakura:2022axo}, we have to deduct $N$ by one.}.

$Z_{\parallel\perp_1}(R)$ is the partition function of the $\parallel\perp_1$ fields
in the background of the $\perp_2$ fields,
\[
Z_{\parallel\perp_1}(R)=\int d\bpsip\cdots d\sigma_{\perp_1} e^{S_{\parallel\perp_1}(R)},
\label{eq:Zpara1}
\]
where $R$ denotes the classical backgrounds of the $\perp_2$ fields, as will be explained below in more detail. 
Here the action $S_{\parallel\perp_1}(R)$ is composed of all the terms which contain the  
$\parallel\perp_1$ fields in \eq{eq:sqnbe} and \eq{eq:s0v}. 
Part of the terms in $S_{\parallel\perp_1}(R)$ contain the $\perp_2$ fields as well. 
For large $N$ these $\perp_2$ fields may well be approximated by their classical values because of the 
large degrees of freedom of the $\perp_2$ fields.
For instance, we perform replacements,
\[
\bar \psi_{\perp_2} \cdot \varphi_{\perp_2} \bpsip \vphip\rightarrow 
\langle \bar \psi_{\perp_2} \cdot \varphi_{\perp_2}\rangle \bpsip \vphip,
\label{eq:replace}
\]
where $\langle \cdot \rangle$ denotes an expectation value.
By doing such replacements we obtain $S_{\parallel\perp_1}(R)$, whose dynamical fields are only the
$\parallel\perp_1$ fields. 

Obtaining the explicit form of $S_{\parallel\perp_1}(R)$ proceeds as follows. 
The quadratic and quartic terms of the $\parallel\perp_1$ fields can be processed in the same manner as 
are performed for $N=2$ in Section~\ref{sec:neq2}, and we obtain $K_{\parallel\perp_1}$ in 
\eq{eq:kineticp1} with \eq{eq:defofa}. Then
the four-interaction terms between the $\parallel\perp_1$ fields and the $\perp_2$ fields, where the 
latter are replaced by their expectation values like in \eq{eq:replace}, 
generate some quadratic terms of the former, which are
explicitly given in \eq{eq:vp12q} of Appendix~\ref{app:intparaonetwo}. Thus we have
\[
S_{\parallel\perp_1}(R)=K_{\parallel \perp_1}+V_{\parallel\perp_1,\perp_2}(R),
\]
whose terms are all quadratic in the $\parallel\perp_1$ fields. 
Then the computation
of the partition function \eq{eq:Zpara1} is just a computation of a determinant, and we obtain
\[
Z_{\parallel\perp_1}(R)=\sqrt{\pi} \int dg_1 dg_2 dg_3\, e^{-g_1^2-g_2^2-g_3^2} \sqrt{\det H},
\label{eq:intexpzp1}
\]
where $H$ is given by 
\[
H=\left (
\begin{array}{cccc}
\epsilon -A_1 R_{22} & a_1 -A_1 R_{12} & 0 & a_2 \\
a_1-A_1 R_{12} & -1-A_1 R_{11} & a_2 &0 \\
0 & a_2 & \epsilon-A_2 R_{22} & a_3-A_2 R_{12} \\
a_2 & 0 & a_3-A_2 R_{12} & -1-A_2 R_{11}
\end{array}
\right),
\label{eq:defofH}
\] 
where $a_i$ are given in \eq{eq:defofa}, $R_{ij}$ are the values of the two point of correlation functions of 
the $\perp_2$ fields explicitly given in \eq{eq:qvalsmall} and \eq{eq:qvallarge}, and 
\[
A_1=\frac{8 \beta v^2 (N-2) }{v^4+12 \alpha \beta},\ A_2=\frac{v^2(N-2)}{3 \alpha}.
\label{eq:Ai}
\]
The derivation of $H$ is given in Appendix~\ref{app:intparaonetwo}.

\section{Comparison with numerical simulations}
\label{sec:comp}
In this section we compare the distributions obtained for the spiked tensor in Sections~\ref{sec:signed} 
and \ref{sec:dist} with Monte Carlo (MC) simulations. 
The method is basically the same as that taken
in the previous works of the author \cite{Sasakura:2022zwc,Sasakura:2022iqd,Sasakura:2022axo}.
Throughout this section we put $\alpha=1/2$ without loss of generality. 
In the MC simulations, all the solutions to the eigenvector equation \eq{eq:egeq} must be computed for 
any randomly sampled $C$ and $\eta$. Since this requires a reliable polynomial equation solver,  
we used Mathematica 13 for the MC simulations.
It computes the solutions to \eq{eq:egeq}, which are generally complex, 
among which we take only the real ones.
To check whether all the solutions are covered, we checked whether
the number of the generally complex solutions to \eq{eq:egeq}
agreed with the number $2^N-1$ of the generally complex eigenvectors proven in \cite{cart},
every time the solutions were computed.
In fact, when $N$ is large, we encountered some
cases that a few solutions were missing. 
However, the missing rates were too small to statistically be relevant for this study. 
For example the missing rate was $\lesssim 10^{-4}$ in the $N=9$ data we use in this paper.
We used a workstation 
which had a Xeon W2295 (3.0GHz, 18 cores), 128GB DDR4 memory, and Ubuntu 20 as OS.

The Monte Carlo simulations were performed by the following procedure.
\begin{itemize}
\item
Randomly sample $C$ and $\eta$. Each $\eta_a$ is randomly sampled by the normal distribution 
with the mean value zero and the standard deviation $\sqrt{2 \beta}$.
Each $C_{abc}$ is randomly sampled by the normal distribution with the mean value zero and the standard 
deviation $1/\sqrt{d_{abc}}$, corresponding to $\alpha=1/2$, 
where $d_{abc}$ is the degeneracy factor defined 
by\footnote{This degeneracy factor is because 
the Gaussian term in \eq{eq:average} is $C_{abc}C_{abc}=\sum_{a\leq b\leq c=1}^N d_{abc} C_{abc}^2$
in terms of the independent components of the symmetric tensor $C$.}
\[
d_{abc}=\left\{
\begin{array}{ll}
1 & \hbox{for } a=b=c,  \\
3 & \hbox{for } a\neq b=c \hbox{ or } b\neq c=a \hbox{ or } c\neq a=b, \\
6 & \hbox{for } a\neq b\neq c \neq a.
\end{array} 
\right.
\label{eq:degen}
\]
\item
As explained above, compute all the complex solutions to the eigenvector equation \eq{eq:egeq}, 
and pick up only the real ones $v^i\ (i=1,2,\cdots, \#{\rm sol}(C,\eta))$. 
\item
Store $\left(|v^i|, v^i\cdot n/|v^i|, {\rm sign}\left( \det M(v^i,Q,C) \right)\right)$ for $i=1,2,\cdots, \#{\rm sol}(C,\eta)$.
\item Repeat the above processes. 
\end{itemize}
By this sampling procedure, we obtain a series of data, 
$\left(|v^h|, v^h\cdot n/|v^h|, {\rm sign}\left( \det M(v^h,Q,C) \right)\right)$ for $h=1,2,\cdots,L$, where 
$L$ denotes the total number of real solutions obtained\footnote{Note that $L$ is generally 
different from $N_{MC}$ below.}. 

To plot the distributions, we classify the data 
into equally spaced bins in $v$ and angle $\theta$ as
\s[
&v-\delta_v/2 < v^h \leq v+\delta_v/2,\\
&\cos(\theta-\delta_\theta/2)<v^h\cdot n/|v^h| \leq \cos(\theta+\delta_\theta/2),\\
&{\rm sign}\left( \det M(v^h,Q,C) \right)=1\hbox{ or }{\rm sign}\left( \det M(v^h,Q,C) \right)=-1,
\label{eq:bincriteria}
\s]  
where $v, \theta$ are the center values of a bin, and $\delta_v,\delta_\theta$ are
the sizes of a bin. We denote the total number of data satisfying \eq{eq:bincriteria} 
as ${\cal N}_{\delta_v,\delta_\theta,+}(v,\theta)$ and ${\cal N}_{\delta_v,\delta_\theta,-}(v,\theta)$
for ${\rm sign}\left( \det M(v^h,Q,C) \right)=1$ and ${\rm sign}\left( \det M(v^h,Q,C) \right)=-1$, respectively. 

Then the distribution of the real eigenvectors from a data is given by
\[
\rho_{MC} (v,\theta;q,\beta)=\frac{1}{N_{MC}\delta_v \delta_\theta} \left( 
{\cal N}_{\delta_v,\delta_\theta,+}(v,\theta)+{\cal N}_{\delta_v,\delta_\theta,-}(v,\theta)
\pm \sqrt{{\cal N}_{\delta_v,\delta_\theta,+}(v,\theta)+{\cal N}_{\delta_v,\delta_\theta,-}(v,\theta)} 
\right),
\label{eq:mcdist}
\]
where $N_{MC}$ denotes the total number of sampling processes in obtaining the data
and the $\pm$ part represents error estimates. 
The signed distribution is given by
\[
\rho_{MC}^{\rm signed} (v,\theta;q,\beta)=\frac{1}{N_{MC}\delta_v \delta_\theta} \left( 
{\cal N}_{\delta_v,\delta_\theta,+}(v,\theta)-{\cal N}_{\delta_v,\delta_\theta,-}(v,\theta)
\pm \sqrt{{\cal N}_{\delta_v,\delta_\theta,+}(v,\theta)+{\cal N}_{\delta_v,\delta_\theta,-}(v,\theta)} 
\right).
\label{eq:mcsigned}
\]

As for the analytical side, since we take only the size $|v|$ and the relative angle $\theta$
as data, the above MC distributions should be compared with 
\s[
\rho_{\rm analy}(v,\theta;q,\beta) dv d\theta =&\int_{|v'|=v,\, v'\cdot n/|v'|=\cos(\theta)}  d^N v'  \rho(v',q,n,\beta)
\\
=&S_{N-2} v^{N-1} \sin^{N-2}(\theta) \, \rho(v,q,n,\beta) dv d\theta, 
\label{eq:distanaly}
\s]
where $S_{N-2}=2 \pi^{(N-1)/2} /\Gamma[(N-1)/2]$ 
is the surface volume of a unit sphere in the $N-1$-dimensional flat space.
Here $\rho(v,q,n,\beta)$ is one of the expressions obtained in Sections~\ref{sec:signed} 
and \ref{sec:dist}, and 
$v$ in the argument of $\rho$ on the righthand side abusively denotes an arbitrary vector $v'$ 
which satisfies $|v'|=v,\, v'\cdot n/|v'|=\cos(\theta)$.
In the following we will compare the Monte Carlo and the analytical results. 

\begin{figure}
\begin{center}
\includegraphics[width=15cm]{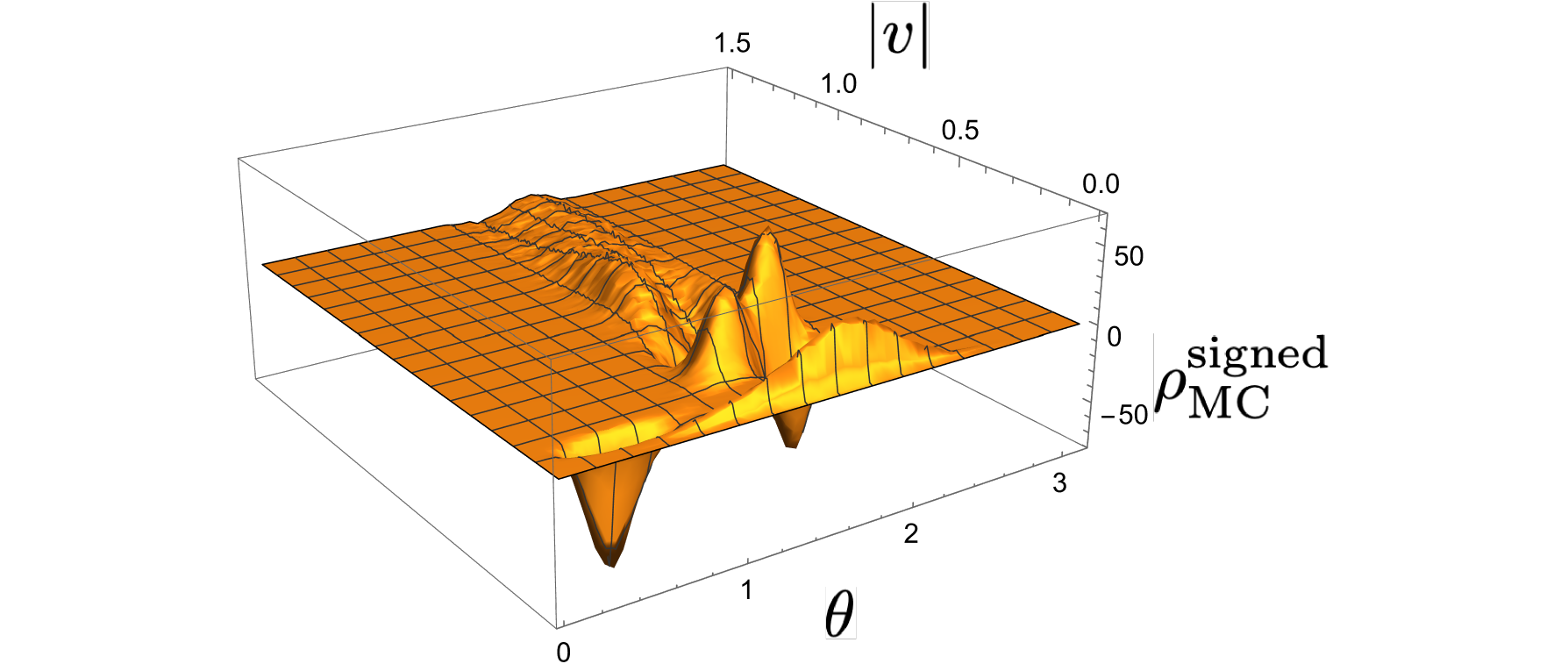}
\caption{The MC signed distribution \eq{eq:mcsigned} is plotted for a data with $N=9,\beta=10^{-4},q=10$
and total sampling number $N_{MC}=4\cdot 10^4$.}
\label{fig:signed}
\end{center}
\end{figure}

\begin{figure}
\begin{center}
\includegraphics[width=7cm]{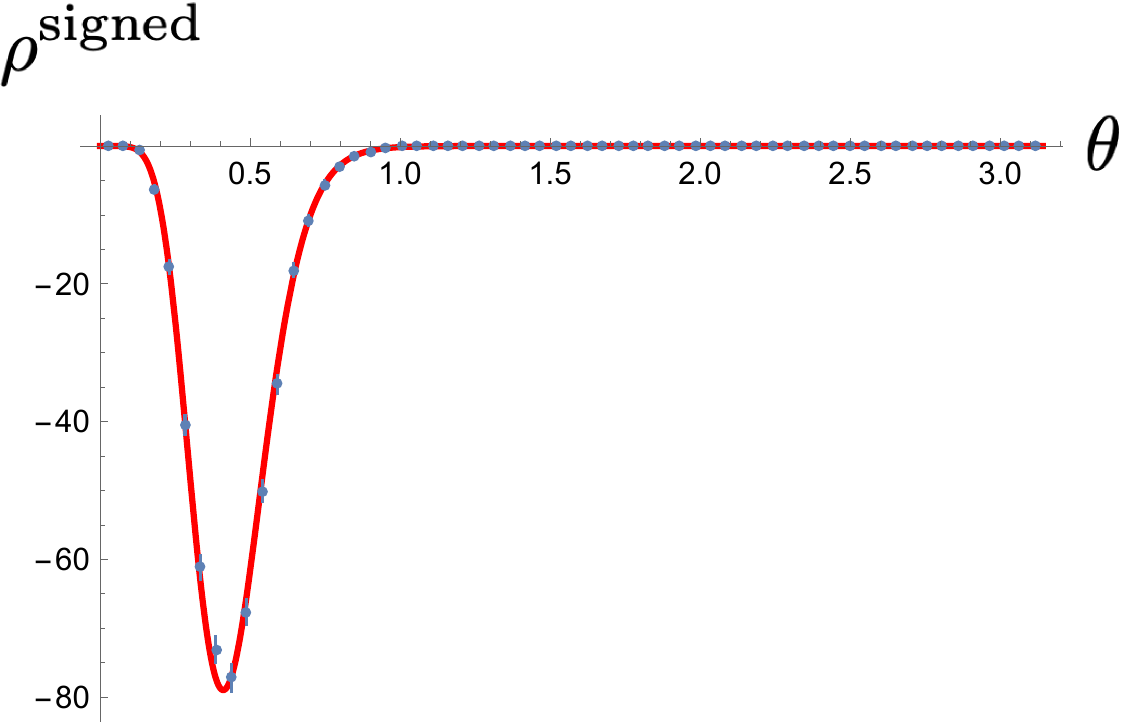}
\hfil
\includegraphics[width=7cm]{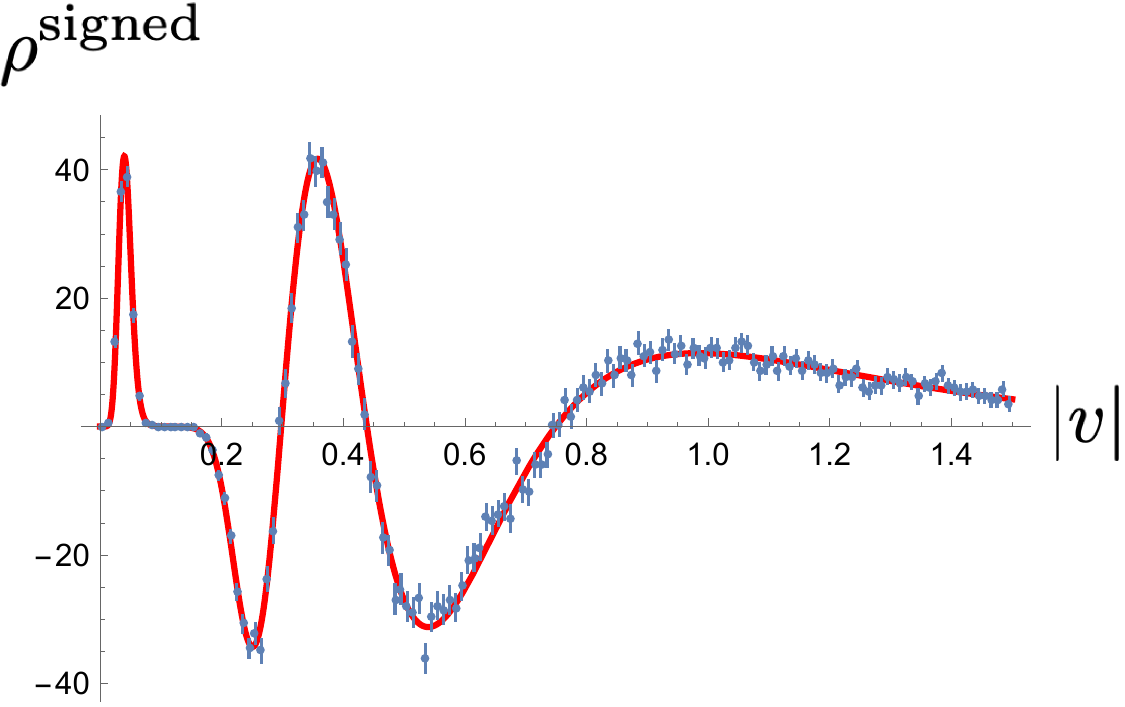}
\caption{The comparison between the analytical and the MC results with the same data as of Figure~\ref{fig:signed}. 
The analytical result is drawn by the solid lines and the MC results are plotted with error bars. 
The comparisons are shown for two example slices in $|v|$ and $\theta$;
the left is at $|v|=0.105$ and the right is at $\theta=\pi/2$. }
\label{fig:signedvth}
\end{center}
\end{figure}

Let us first consider the signed distribution. 
The analytical result is obtained by putting \eq{eq:rhopen} with \eq{eq:exact} into \eq{eq:distanaly}.
Since the analytical result is an exact result, it should agree with the MC result within errors. 
In Figure~\ref{fig:signed}, we plot the MC result \eq{eq:mcsigned} 
for $N=9,\beta=10^{-4},q=10$ with $N_{MC}=4\cdot 10^4$.
As examples, the analytical and MC results are compared at two slices,
one at $|v|=0.105$ and the other at $\theta=\pi/2$ in the two slots of Figure~\ref{fig:signedvth}.
They agree quite well within error estimates, 
supporting the validities of both the analytical and the MC computations.

As in Figure~\ref{fig:signed} and the left slot of Figure~\ref{fig:signedvth}, 
an evident negative peak can be observed around $|v|\sim 0.1$ and 
$\theta\sim 0.5$. This peak approximately corresponds to an eigenvector $q^{-1} n_a$ of the 
background tensor $Q_{abc}=q\, n_a n_b n_c$.
In fact, the location satisfies $|v|\sim q^{-1}$,  while the angle is not strictly $\theta=0$. 
The reason is that the volume factor in \eq{eq:distanaly} contains $\sin^{N-2}(\theta)$,
and pushes the peak away from $\theta=0$.
Because of the same reason,  the other major structures are concentrated around $\theta=\pi/2$ 
in Figure~\ref{fig:signed}. 
A large-$N$ limit which effectively vanishes this volume effect will be discussed in Section~\ref{sec:largen}.

In Figure~\ref{fig:signed} and the right slot of Figure~\ref{fig:signedvth} 
one can also see a peak around $|v|\sim 0.04, \theta\sim\pi/2$. 
This peak corresponds to the trivial eigenvector $v=0$.
Because of $\beta> 0$ the distribution broadens around $|v|\sim 0$, and the volume factor $v^{N-1}$ in
\eq{eq:distanaly} pushes the peak away from $|v|=0$. 

\begin{figure}
\begin{center}
\includegraphics[width=10cm]{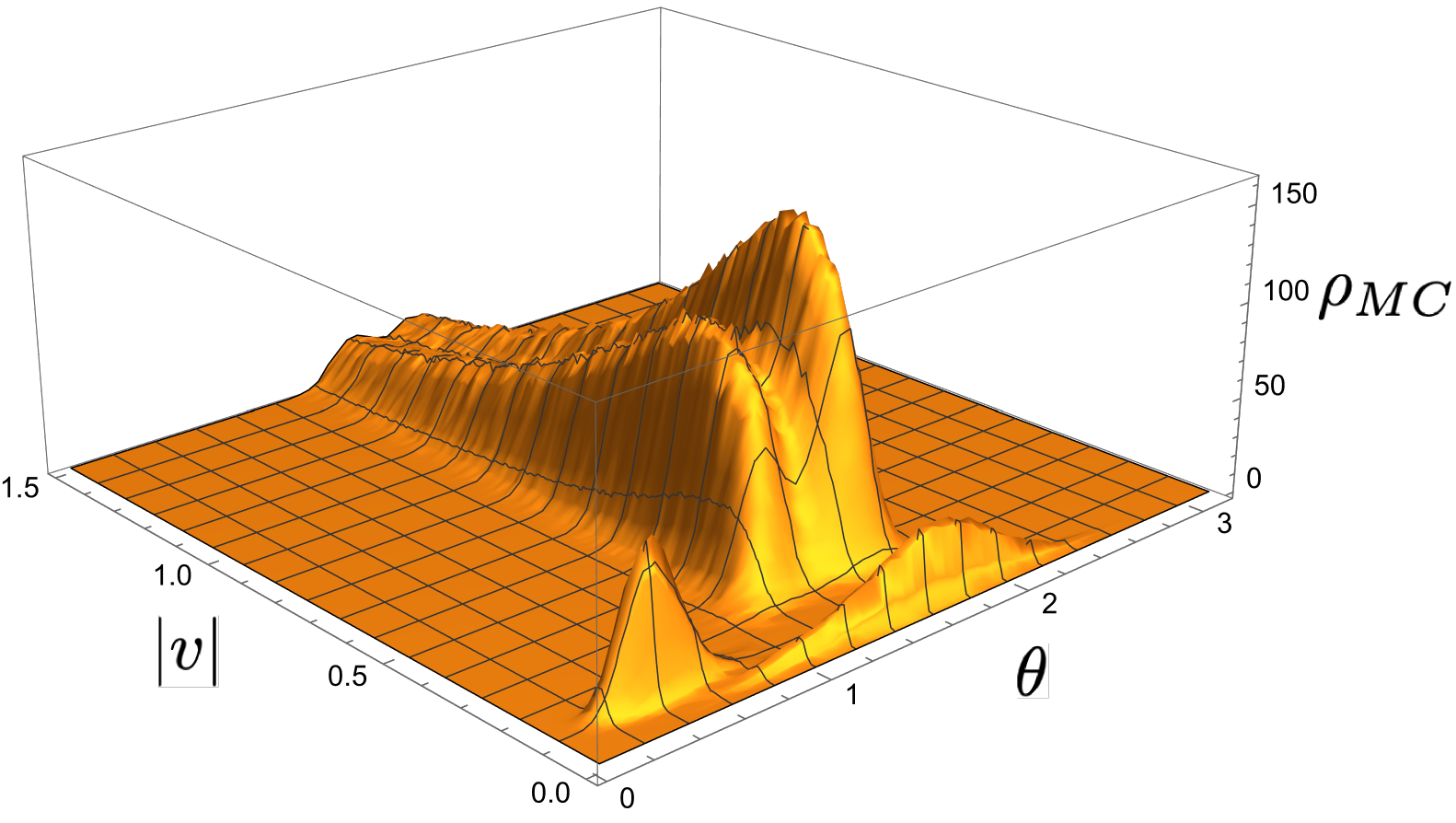}
\caption{The MC distribution \eq{eq:mcdist} is plotted for the same data
as used in Figure~\ref{fig:signed} for $N=9,\beta=10^{-4},q=10$ and $N_{MC}=4\cdot 10^4$. }
\label{fig:dist}
\end{center}
\end{figure}

\begin{figure}
\begin{center}
\includegraphics[width=7cm]{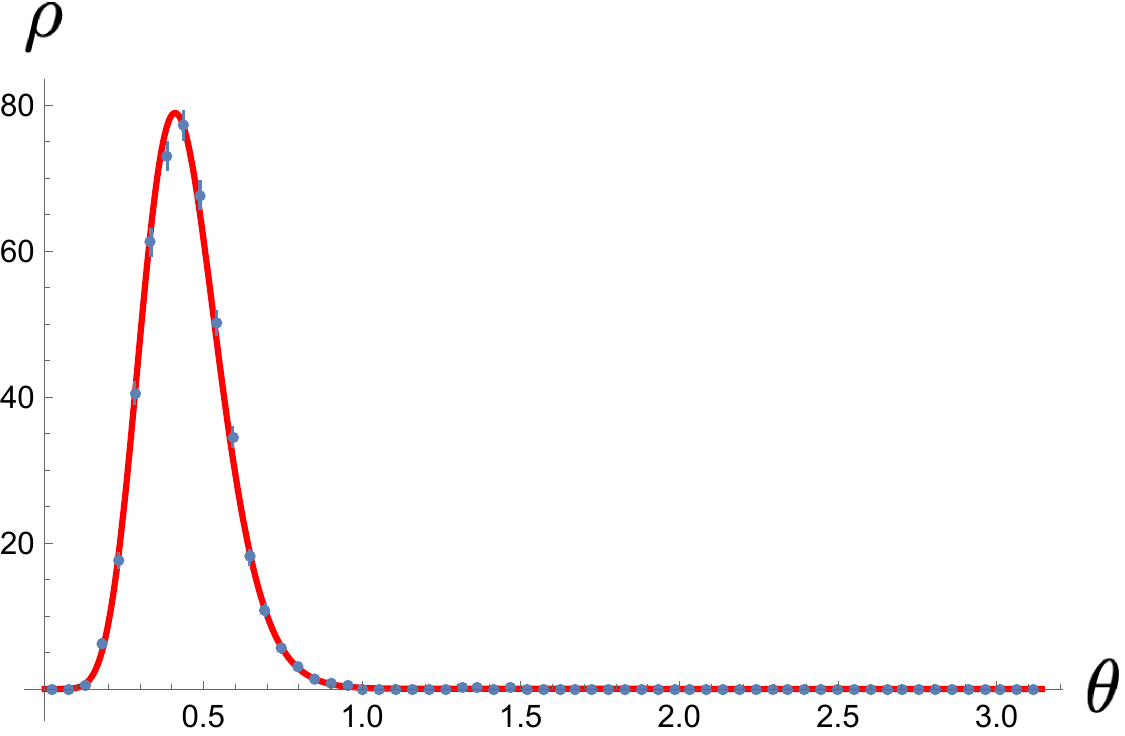}
\hfil
\includegraphics[width=7cm]{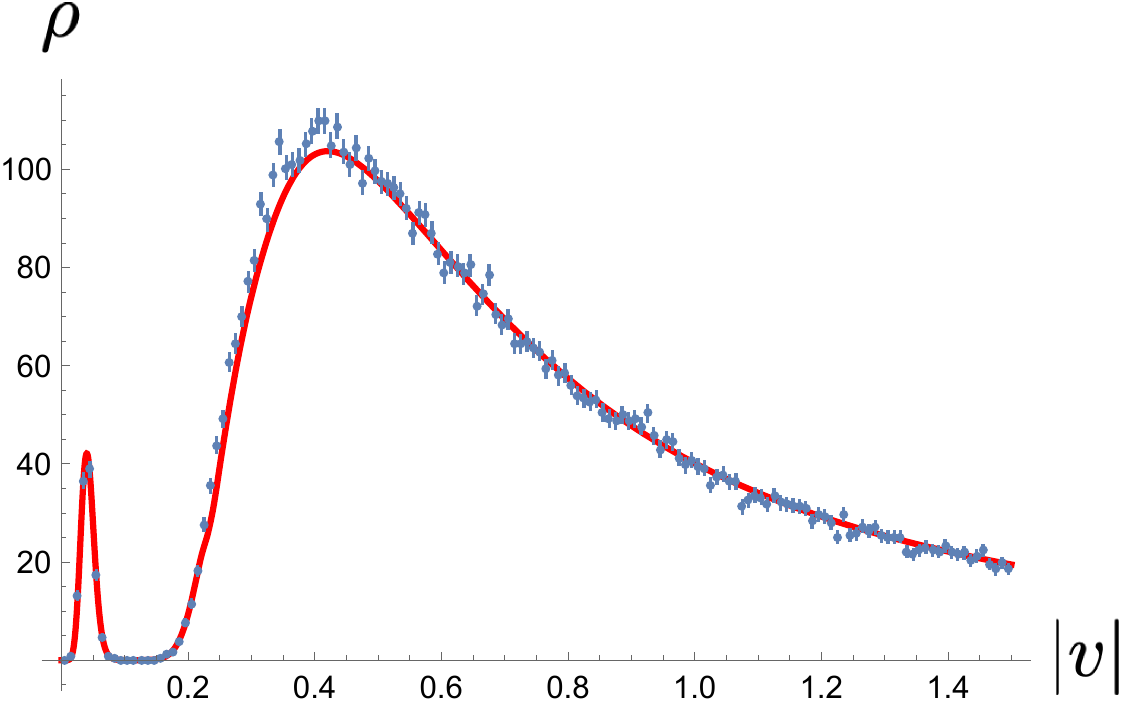}
\caption{The MC results are plotted with error bars for the same data as in Figure~\ref{fig:dist}.
The analytic result is drawn by the solid lines. 
The left slot is of the slice at $|v|=0.105$, and the right at $\theta=\pi/2$.} 
\label{fig:distvth}
\end{center}
\end{figure}

In Figure~\ref{fig:dist} the MC distribution \eq{eq:mcdist} is shown for the same data.
Except for the signs, the characters of the distribution are more or less similar to the signed case. 
On the other hand, the analytic result for this case has the difference that the partition function $Z$ 
in \eq{eq:rhodist} is computed by the approximation \eq{eq:zapp}, while it was exact for the signed case.
The exact expression of $Z_{\perp_2}$ can be taken from the previous result in \cite{Sasakura:2022axo}, 
which is explicitly given in Appendix~\ref{app:exactz1}. 
As for $Z_{\parallel \perp_1}$, by numerically integrating \eq{eq:intexpzp1} on a grid of points in 
$|v|$ and $\theta$, an interpolation function of $Z_{\parallel \perp_1}$ is computed and used.
In Figure~\ref{fig:distvth} the analytic and the MC results are compared. 
The agreement is fairly satisfactory except for some slight systematic deviations around a peak.  

\section{Large-$N$ limit}
\label{sec:largen}
In this section we will take large-$N$ limits of the distribution obtained in Section~\ref{sec:rankone} 
for a spiked tensor. 
We will particularly pay attention to the parameter region where the peak corresponding
to the background $Q$ can been seen in the eigenvector distribution. 
We will consider two large-$N$ limits. 
In one large-$N$ limit, we will derive the result 
that a peak can be well identified with $Q$ for the parameter region, 
$\alpha q^2 /N \gtrsim 0.6  , \beta q^2 N \lesssim 0.1$.
In particular for $\beta q^2 =0$, we will find the threshold value to be $0.66<(\alpha q^2 /N)_c<0.67$, 
which agrees with Proposition 2 of \cite{AMN}.
However this peak is always smaller than the other peak(s) at $\np=0$ 
and therefore relatively vanishes in the strict large-$N$ limit. 
In the other scaling limit,  $\alpha q^2 \sim N^\gamma, \beta  q^2 \sim N^{-\gamma}$ with $\gamma>1$,
the peak remains in the strict large-$N$ limit. 

We want to consider large-$N$ limits which keep both the parameters $Q$ and $\beta$ relevant.
As was discussed in Section~\ref{sec:comp}, the volume factor $\sin^{N-2}\theta$ in \eq{eq:distanaly}  
suppresses the peak of the eigenvector $q^{-1} n$ of the background tensor $Q$, and 
this suppression becomes stronger as $N$ becomes larger.
Therefore, to obtain an interesting large-$N$ limit, the parameters must be scaled so as to 
compete with $\sin^{N-2}\theta \sim e^{N \log(\sin\theta)}$.
A large-$N$ scaling which makes the exponential factor in \eq{eq:rhodistspiked} in this order
is given by 
\[
\alpha=\frac{N \tilde \alpha}{q^2},\ \beta=\frac{\tilde \beta}{N q^2},\ v=\frac{\tilde v}{q},
\label{eq:scaling}
\]
where $\tal,\tbe$ are kept finite. Here the factors of $q$ are to absorb the dependence on $q$ 
from the formulas below.

Let us discuss the large-$N$ limit of $Z=Z_{\perp_2} Z_{\parallel\perp_1}$ in Section~\ref{subsec:largen}. 
The large-$N$ limit of $Z_{\perp_2}$ was determined in \cite{Sasakura:2022iqd}, and it is given by
\[
Z^{N=\infty}_{\perp_2}\sim {\rm const.} \,e^{N S^{\infty}_{\perp_2}},
\]
where\footnote{For simplicity, 
$S^{\infty}_{\perp_2}$ is shifted by an irrelevant constant from the corresponding expression with $R=1/2$ 
in \cite{Sasakura:2022iqd}.}
\[
S^{\infty}_{\perp_2}(x)=\left\{
\begin{array}{ll}
\log 2+ \log(x) +\frac{1-\sqrt{1-4 x}}{4x}- \log\left(1-\sqrt{1-4 x}\right)& \hbox{ for } 0<x<\frac{1}{4}, \\
 \frac{1}{4x} +\frac{1}{2}  \log(x)& \hbox{ for } \frac{1}{4}<x,
\end{array}
\right.
\label{eq:seffzero}
\]
with\footnote{$N$ must be deducted by one, when we take a result from \cite{Sasakura:2022iqd}. 
See a footnote below \eq{eq:fermitwo}.}
$x=(N-2) v^2/(3 \alpha) \sim \tilde v^2 /(3 \tilde \alpha)$.
As for $Z_{\parallel\perp_1}$, one can easily see that the limit of \eq{eq:intexpzp1} is just given by 
dropping the terms dependent on $g_i$ in \eq{eq:defofa}, while the $N$-dependencies of $A_i$ in \eq{eq:Ai}
and $R_{ij}$ in \eq{eq:qvalsmall} and \eq{eq:qvallarge} drop out. Therefore $H$ does not 
depend on $g_i$ and we get
\[
Z_{\parallel \perp_1}^{N=\infty}=\pi^2 \left. \sqrt{\det H} \right|_{g_i=0},
\]
which has no relevant effects to the formula below for the large $N$ limit.

By collecting the results above and using \eq{eq:distanaly} and \eq{eq:rhodistspiked},
we obtain
\s[
S_{\infty}(\tilde v,\theta) =&
\lim_{N\rightarrow \infty} \frac{1}{N} \log \rho_{\rm analy}  \\
=&\hbox{const.} + S_{\perp_2}^\infty + \log \tv  +\log(\nt) -\frac{1}{2} \log(\tv^4+12 \tal \tbe) \\ &
+\frac{-\tal \tv^2+2 \tal  \tv^3 \np^3-\tal  \tv^4 \np^6}{\tv^4+4 \tal \tbe} -\frac{3 \tal  \tv^4 \np^4 \nt^2 }{\tv^4+12 \tal \tbe},
\label{eq:infact}
\s]
where $n_\parallel=\cos \theta$ ($n_\perp=\sin \theta$), and \hbox{const.} is the part not dependent 
on $\tv$ or $\theta$.

\begin{figure} 
\begin{center}
\includegraphics[width=8cm]{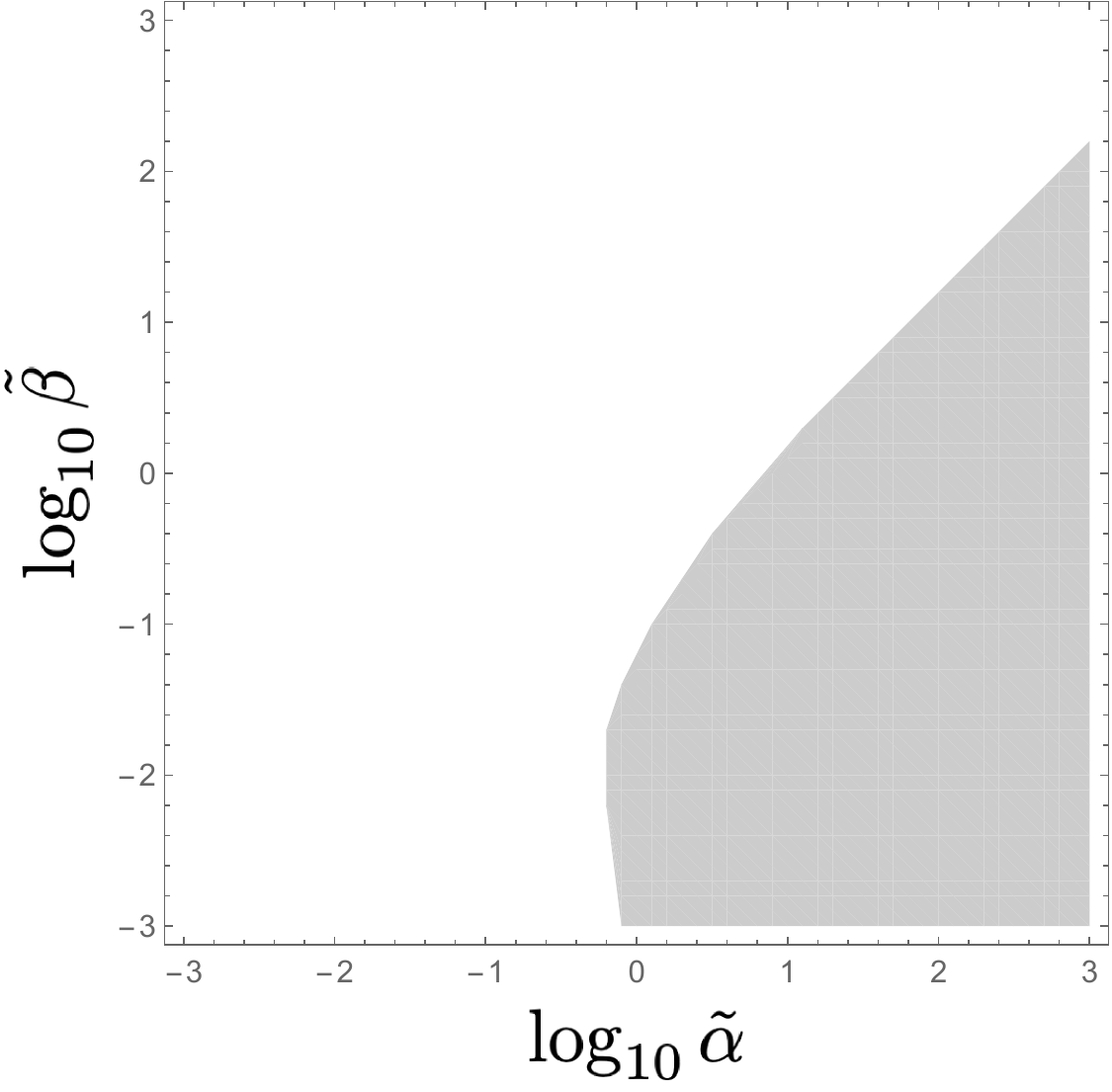}
\caption{In the shaded region of the parameters, 
the eigenvector distribution has a peak of $S_\infty$ corresponding to the eigenvector of $Q$.}
\label{fig:phase}
\end{center}
\end{figure}

\begin{figure}
\begin{center}
\includegraphics[width=7cm]{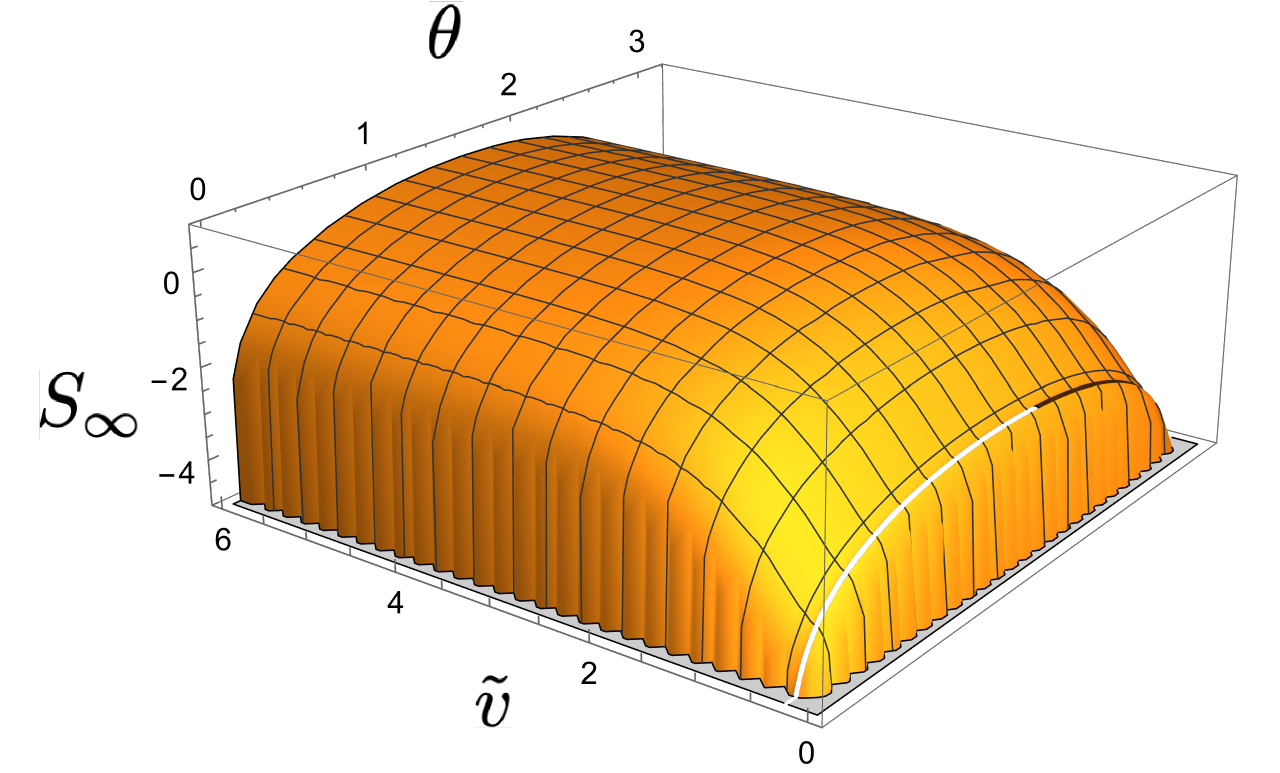}
\hfil
\includegraphics[width=7cm]{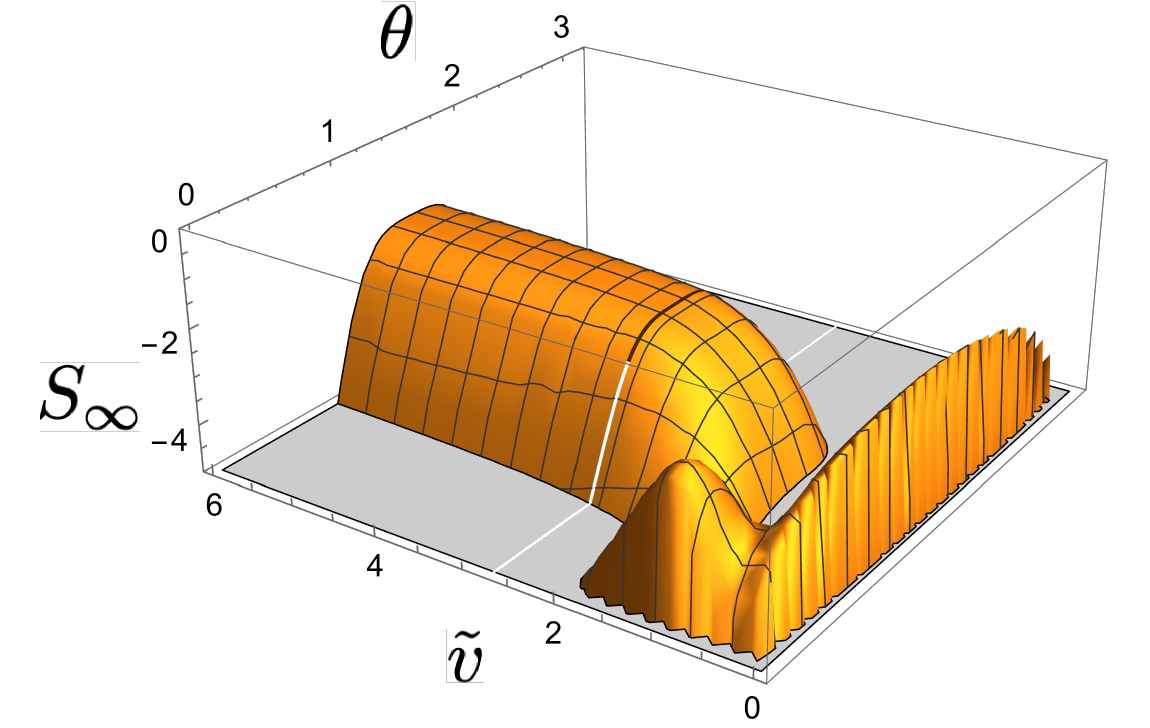}
\caption{In the left slot, $S_{\infty}$ ({const.} being ignored) is plotted for $\log_{10}\tal=-1,\log_{10}\tbe=1$, which
is in the unshaded region of Figure~\ref{fig:phase}. The right slot is 
for $\log_{10}\tal=1,\log_{10}\tbe=-2$ in the shaded region. In the latter case, a peak near $\tv\sim1,\theta\sim0$
corresponding to the eigenvector of $Q$ can be found.
The tiny gaps in the plots are not essential; They seem to be caused by the drawing program
(Mathematica) avoiding the singularity at $x=1/4$ in \eq{eq:seffzero}, where the function is continuous
but its first derivative is discrete.}
\label{fig:ab}
\end{center}
\end{figure}

\begin{figure} 
\begin{center}
\includegraphics[width=8cm]{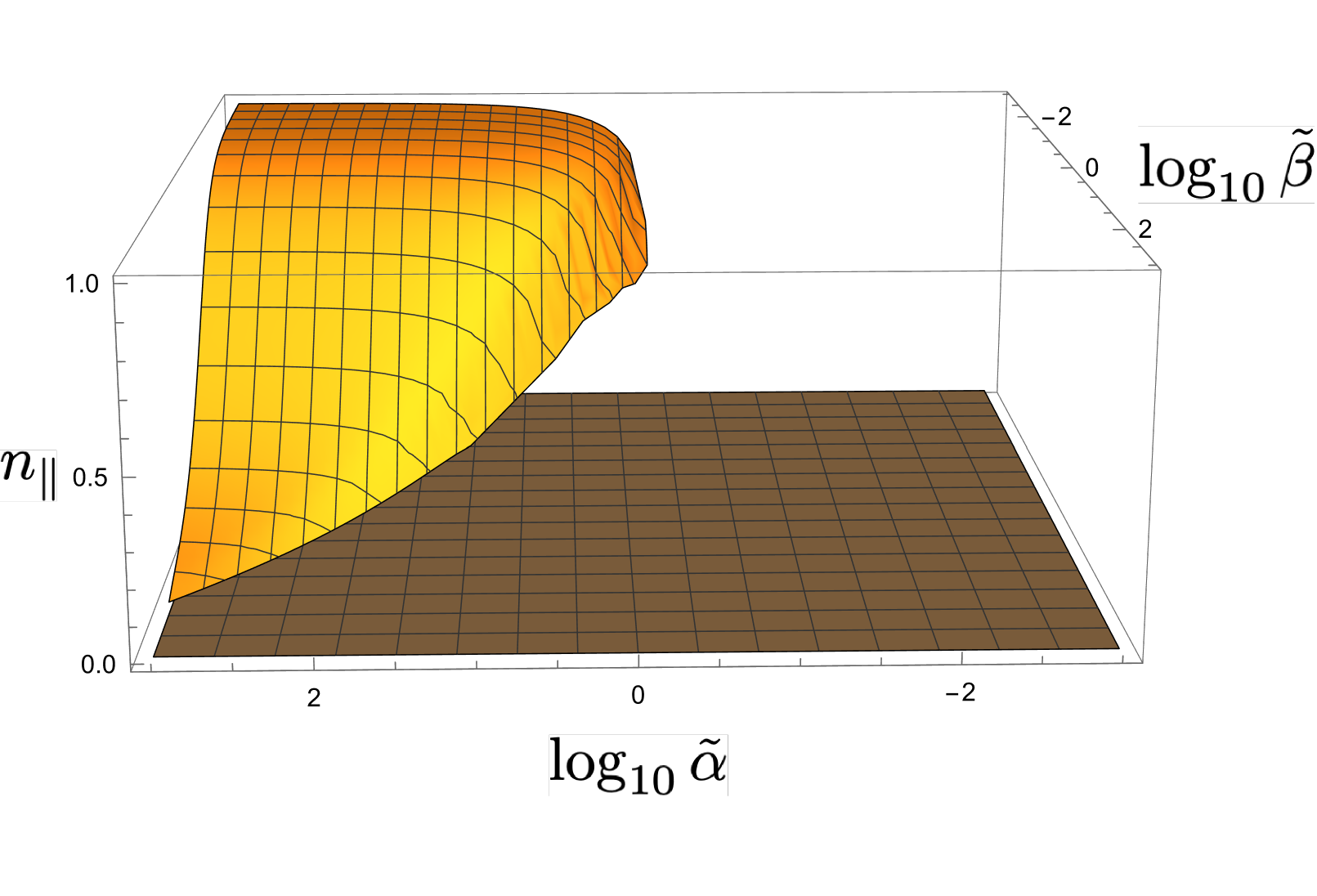}
\hfil
\includegraphics[width=7.7cm]{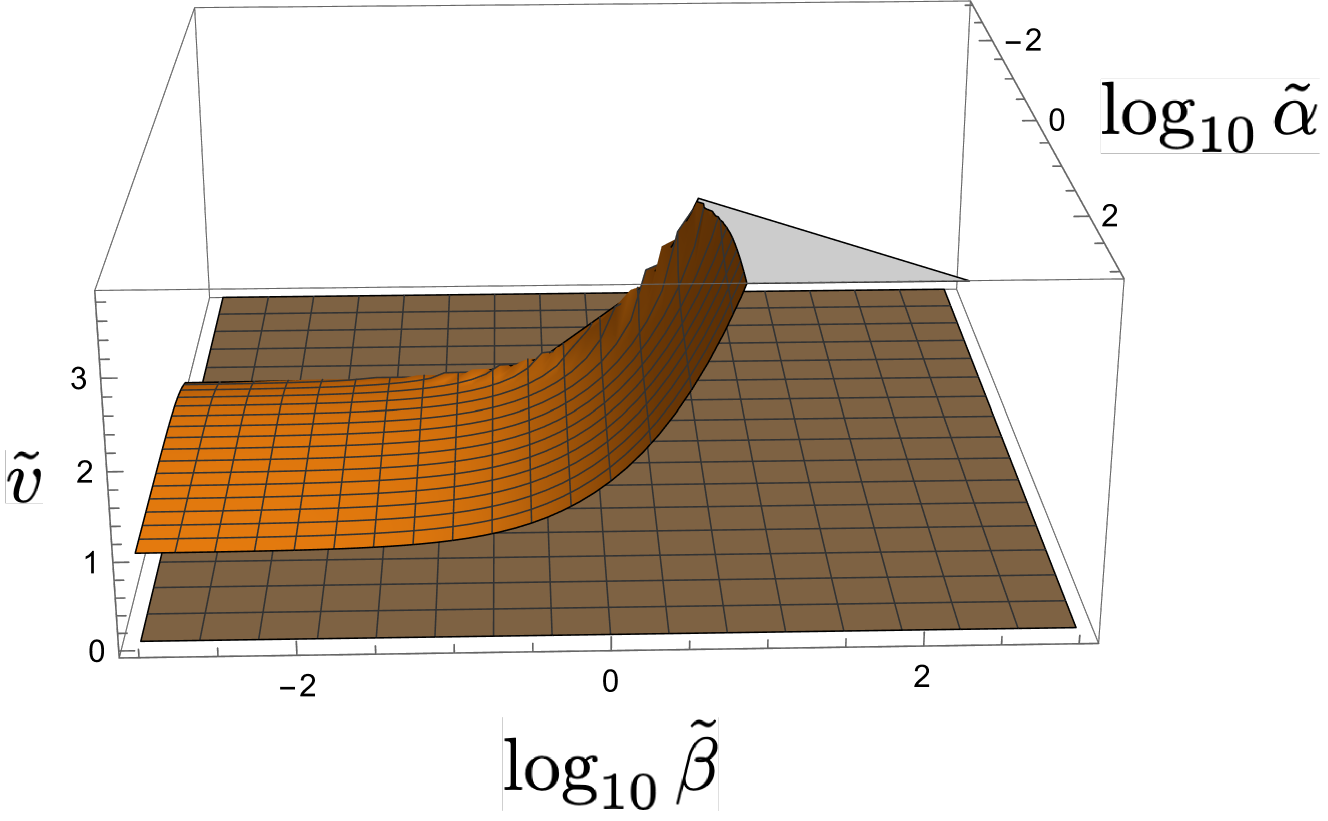}
\caption{The values of $\np$(left) and $\tv$(right) of the peak with $n_\parallel>0$, 
corresponding to the eigenvector of $Q$, are plotted. Identification of this peak with $Q$ 
can well be done in the region  $\log_{10} \tilde \alpha \gtrsim -0.2$ and $\log_{10} \tilde \beta \lesssim -1$.}
\label{fig:npv}
\end{center}
\end{figure}

It is interesting to study the profile of $S_{\infty}(\tv,\theta)$ in the $\tv$ and $\theta$ plane
for various values of $\tilde \alpha, \tilde \beta$. We have 
numerically studied it for the parameter region 
$10^{-3}\leq \tilde \alpha \leq 10^3, 10^{-3}\leq \tilde \beta \leq 10^3$. 
In the unshaded region of Figure~\ref{fig:phase}, the peak(s) exist only along $n_\parallel=0$, as 
is shown in the left slot of Figure~\ref{fig:ab} as an example. In the shaded region, in addition to the peak(s) 
at $n_\parallel=0$, there exists also a peak which has non-zero $n_\parallel$.
This peak corresponds to the eigenvector $q^{-1} n$ of the background tensor $Q$, 
as is shown in the right slot of Figure~\ref{fig:ab} as an example.
In Figure~\ref{fig:npv}, the values of $n_\parallel$ and $\tilde v$ are plotted for the latter peak. 
The location can be well identified with $q^{-1} n$,
if the values take $n_\parallel\sim 1$ and $\tilde v\sim 1$.  As can be seen in the plots, this occurs
in the region, $\log_{10} \tilde \alpha \gtrsim -0.2$ and $\log_{10} \tilde \beta \lesssim -1$.  
This is the parameter region in which the background tensor $Q$ can be detected well.  

\begin{figure}
\begin{center}
\includegraphics[width=7cm]{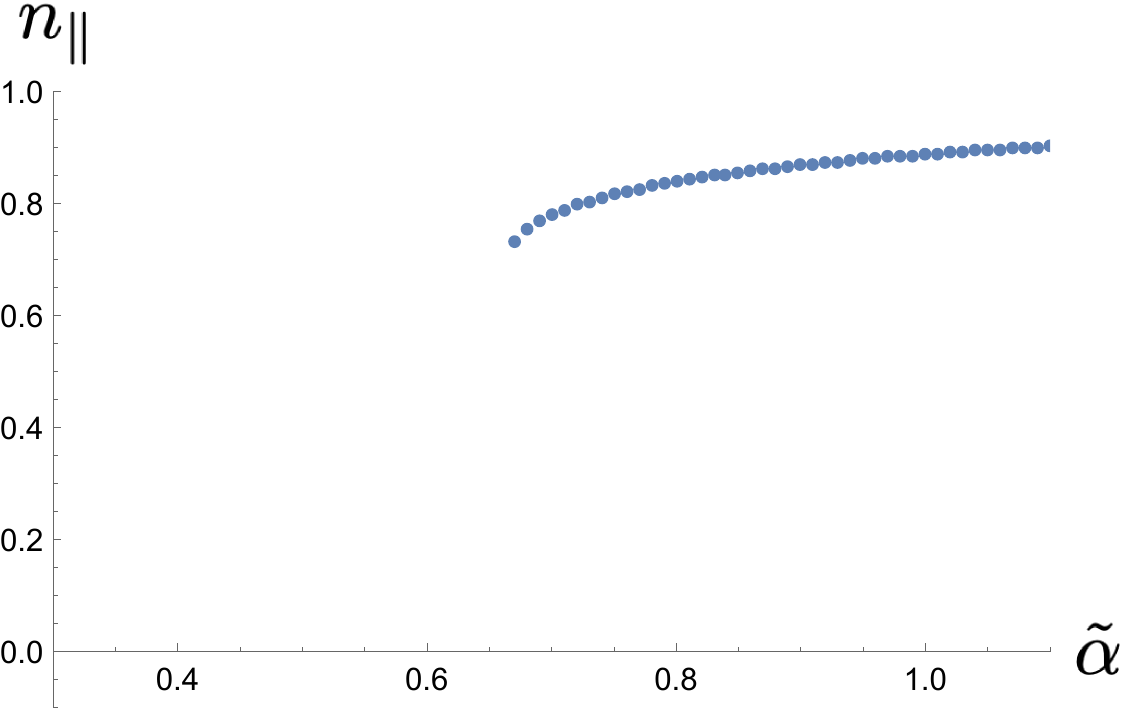}
\hfil
\includegraphics[width=7cm]{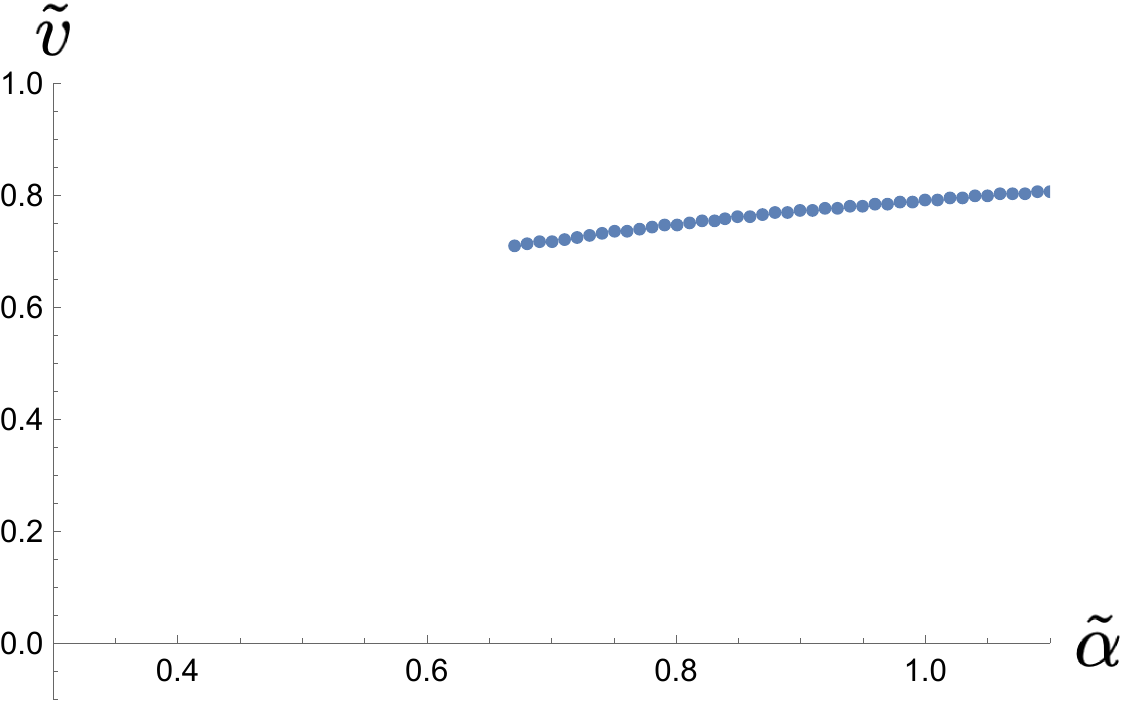}
\caption{The values of $\np$ and $\tv$ of the peak with $\np>0$ for $\tbe=0$. 
The threshold value is $0.66<\tal_c < 0.67$.}
\label{fig:alnv}
\end{center}
\end{figure}

It is interesting to compare this detectable region with a result of \cite{AMN}. 
As can be seen in Figure~\ref{fig:phase}, the shaded region has 
an edge around $\log_{10} \tal \sim -0.2$, namely, $\tal \sim 0.63$, independent of $\tbe$
for $\log_{10} \tbe\lesssim -1$. 
To see the threshold value more precisely for $\tbe=0$, 
we plot $n_\parallel$ and $\tv$ of the peak with $n_\parallel >0$
in Figure~\ref{fig:alnv}. We find that the peak does not exist at $\tilde \alpha\leq 0.66$, 
but exists at $\tilde \alpha\geq 0 .67$ with $\np \gtrsim 0.7$.  
On the other hand, as explained in Appendix~\ref{app:amn},  Proposition 2 of \cite{AMN} 
states that the threshold value is $\tal=2/3$, which indeed agrees with our value.

We numerically observed that a peak at $\np=0$ always take the largest value 
of $S_\infty$ at least in the parameter region of $\tilde \alpha, \tilde \beta$ we have studied above. 
This means that, because $\rho \sim e^{N S_\infty}$,
the peak corresponding to $Q$ will effectively be invisible compared to the peak(s) at $\np=0$ 
in the strict large-$N$ limit.
Therefore in the strict large-$N$ limit, $Q$, namely a ``signal'', cannot be detected by solving 
the eigenvector equation \eq{eq:egeq}.  

\begin{figure}
\begin{center}
\includegraphics[width=7cm]{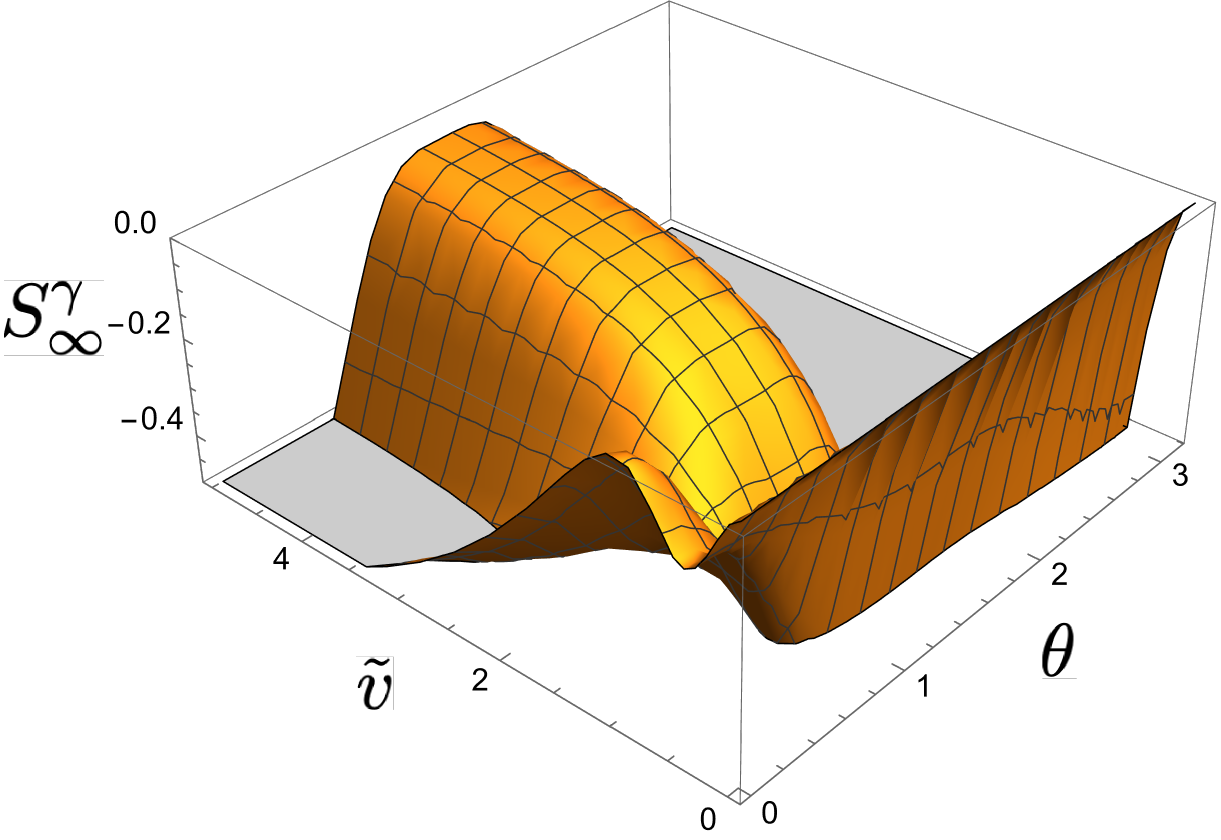}
\caption{$S_\infty^\gamma$ is plotted 
for $\log_{10} \tilde \alpha =0$ and $\log_{10} \tilde \beta = -1$
as an example. There is a peak corresponding to the eigenvector of the background tensor $Q$.}
\label{fig:sgam}
\end{center}
\end{figure}

The main reason for the above difficulty of detection comes from the strong effect of the volume factor
$\sin^{N-2}\theta$ in \eq{eq:distanaly}, which enhances the region $\np\sim 0$ so strongly.
Therefore an obvious way to solve this difficulty is to consider another scaling limit which
overwhelms the volume factor. An example is given by 
\[
\alpha=\frac{N^\gamma \tilde \alpha}{q^2},\ \beta=\frac{\tilde \beta}{N^\gamma q^2},\ v=\frac{\tilde v}{q},
\ \gamma>1.
\label{eq:scaling2}
\]
In this limit, $x=(N-2)v^2/(3 \alpha)\sim N^{-\gamma+1}\rightarrow 0$ in the large-$N$ limit, 
so therefore \eq{eq:seffzero} becomes a constant, meaning that $Z_{\perp_2}$ is a free theory
independent of $v$.
As for $Z_{\parallel \perp_1}$, $A_i\rightarrow 0$ and $R_{ij}$ approaches finite values, 
so $Z_{\parallel \perp_1}$ 
is again a finite quantity.  
Therefore from \eq{eq:rhodistspiked}, the major contribution comes only from the exponent, and we obtain
\s[
S_{\infty}^\gamma (\tilde v,\theta) =&
\lim_{N \rightarrow \infty} \frac{1}{N^\gamma} \log \rho_{\rm analy}  \\
=&\frac{-\tal \tv^2+2 \tal  \tv^3 \np^3-\tal  \tv^4 \np^6}{\tv^4+4 \tal \tbe} -\frac{3 \tal  \tv^4 \np^4 \nt^2 }{\tv^4+12 \tal \tbe}.
\label{eq:pact}
\s]
As is shown in Appendix~\ref{app:maxs}, it is straightforward 
to prove that the maximum value of $S_{\infty}^\gamma$ is $0$, and
this occurs only at three locations: (i) $\tilde v=0$, (ii) $\tilde v \rightarrow \infty,n_\parallel=0$, 
(iii) $\tilde v=1, n_\parallel=1$. The last location corresponds to the background $Q$. 
An example of $S^\gamma_\infty$ is shown in Figure~\ref{fig:sgam}.
Since the eigenvector distribution is given by $\rho\sim e^{N^\gamma S_\infty^\gamma}$, 
there remains only the three locations above in the strict large-$N$ limit.
This means that, in the limit, a finite eigenvector ($v\neq 0,\infty$) is surely that of the background $Q$.  

\section{Summary and future prospects}
In this paper we have studied the real eigenvector distributions of real symmetric order-three 
Gaussian random tensors
in the case that the random tensors have non-zero mean value backgrounds and the eigenvector equations
have Gaussian random deviations. This is an extension of the previous studies 
\cite{Sasakura:2022zwc,Sasakura:2022iqd,Sasakura:2022axo}, which have no such mean values 
or deviations.
We have derived the quantum field theories with quartic interactions whose partition functions give
the distributions. 
For the background tensor being rank-one (a spiked tensor case) in particular, we have explicitly derived the 
distributions by computing the partition functions exactly or approximately. 
We have obtained good agreement between the analytical results and Monte Carlo simulations. 
We have derived the scaling and range of parameters for the background tensor to be 
detectable in the distributions in the large-$N$ limit. Our threshold value has agreed with that of \cite{AMN}. 

The quantum field theories we have derived in this paper are much more complicated than those
in the previous studies \cite{Sasakura:2022zwc,Sasakura:2022iqd,Sasakura:2022axo} 
due to the presence of the backgrounds and the deviations. 
Nonetheless,  we have obtained some exact expressions for the signed distributions, 
and have also derived some approximate expressions of the (authentic) distributions,
which agree very well with the Monte Carlo results. 
This success can be ascribed to the quantum field theoretical expressions, to which we can apply 
various well-developed techniques and knowledge of quantum field theories.
The results of this paper strengthen our belief that the quantum field theoretical procedure for computing
distributions of quantities in random tensors is general, powerful, and intuitive.

As far as random tensors are Gaussian, it is in principle straightforward to extend the quantum field theoretical 
procedure to some other problems in random tensors; distributions of complex eigenvectors/values, 
tensor rank decompositions, correlations among eigenvectors, etc.
Although derived quantum field theories with quartic interactions may become quite complicated, it will
always be possible to find ways to, exactly or approximately, 
compute the partition functions by quantum field theoretical techniques, knowledge and intuition. 
These studies will enrich fundamental knowledge about random tensors, which will eventually be applied 
in various subjects in future studies.

Tensor models have emerged from discrete approaches to quantum 
gravity \cite{Ambjorn:1990ge,Sasakura:1990fs,Godfrey:1990dt,Gurau:2009tw}, and are also taking 
active part in more recent approaches, such as in the AdS/CFT correspondence \cite{Witten:2016iux}. 
A question of the author's interest is whether there exists an
analogous phenomenon in tensor models as the 
Gross-Witten-Wadia transition \cite{Gross:1980he,Wadia:1980cp}. In fact 
there are some indications that similar transitions exist in the context of a discrete model of quantum 
gravity \cite{Kawano:2021vqc,Sasakura:2022aru}. 
We hope the knowledge about random tensors enriched along the line of our studies 
will give some insights into quantum gravity in the future.

\section*{Acknowledgment}
The author is supported in part by JSPS KAKENHI Grant No.19K03825. 
He would like to thank M.~Akyazi,  N.~Delporte, O.~Evnin, R.~Gurau, L.~Lionni, Z.~Mirzaiyan,
and R.~Toriumi for some stimulating discussions.


%

\vspace{0.2cm}
\noindent

\let\doi\relax


\appendix

\section{Derivation of \eq{eq:rhosignfinal}}
\label{app:dels}
From \eq{eq:defofdsgn}, the parallel/transverse parts of $D^{\rm signed}$ are given by
\s[
D^{\rm signed}_\parallel&=\frac{v^3}{\alpha} \bpsip \psip, \\
D^{\rm signed}_\perp &= \frac{v^3}{3 \alpha} \left( \bpsit \psip +\bpsip \psi_\perp\right).
\label{eq:ds}
\s]
By putting \eq{eq:ds} into \eq{eq:subdels}, we obtain
\s[
\delta S^{\rm signed}_{\lambda}-\delta S^{\rm signed}_{\lambda}(Q=\beta=0)&=\frac{1}{2} \log b_\parallel +\frac{N-1}{2} \log b_\perp -\frac{\alpha(b_\parallel-1)}{v^2} +\frac{2 \alpha b_\parallel D^Q_\parallel}{v^3} -\frac{\alpha b_\parallel (D^Q_\parallel)^2}{v^4}  \\ 
&\ \ -\frac{3 \alpha b_\perp D^Q_\perp \cdot D^Q_\perp}{v^4} 
-2 \left( b_\parallel-1-\frac{b_\parallel D^Q_\parallel}{v} \right) \bpsip \psip 
\\ &\ \ +\frac{2 b_\perp}{v}  D^Q_\perp \cdot \left( \bpsit \psip +\bpsip \psi_\perp \right)
+\frac{2 v^2(b_\perp-1)}{3 \alpha} \bpsip \psip \bpsit \cdot \psi_\perp. 
\label{eq:delsbar}
\s]
Adding this and the last term of \eq{eq:snew} to \eq{eq:sbarqbzero}, one obtains \eq{eq:rhosignfinal} with \eq{eq:ssignfinal}.

\section{Derivation of \eq{eq:exact}}
\label{app:delofU}

Let us parametrize \eq{eq:spen} as follows:
\s[
S_{b,d,k}&=b_1 \bpsip \psip +b_2 \left(\bpsip \psione+\bpsione \psip
\right)+b_3 \bpsione \psione +k \bpsitwo \psitwo  \\
&\ \ + d_1 \left( \bpsione \psione +\bpsitwo \psitwo\right) \bpsip \psip+d_2 \left( \bar \psi _1 \psione +\bpsitwo\psitwo \right)^2.
\s]
Then by explicitly performing the fermion integrations for $\parallel$ and $\perp_1$ directions, we obtain 
\s[
&\int d\bar \psi d\psi\, e^{S_{b,d,k}}\\
&=\int d\bpsitwo d\psitwo 
\left( d_1+b_1 b_3-b_2^2+(2 b_1 d_2 +b_3 d_1) \bpsitwo \cdot \psitwo 
+2 d_1 d_2 \left( \bpsitwo \cdot \psitwo \right)^2 \right) e^{k \bpsitwo \cdot \psitwo+d_2 \left(\bpsitwo\cdot \psitwo \right)^2} \\
&=
\left( d_1+b_1 b_3-b_2^2+(2 b_1 d_2 +b_3 d_1) \frac{\partial}{\partial k }
+2 d_1 d_2 \frac{\partial^2}{\partial k^2 } \right) 
\int d\bpsitwo d\psitwo \, e^{k \bpsitwo \cdot \psitwo+d_2 \left(\bpsitwo\cdot \psitwo \right)^2}. 
\s]
Now the last fermion integration can be computed as
\s[
\int d\bpsitwo d\psitwo \, e^{k\bpsitwo \cdot \psitwo+d_2 \left(\bpsitwo\cdot \psitwo \right)^2}&
 = \sum_{n=0}^\infty
 \frac{d_2^n}{n!} \left(\bpsitwo\cdot \psitwo \right)^{2n} \int d\bpsitwo d\psitwo \,
 e^{k\bpsitwo \cdot \psitwo}
 \\
&=\sum_{n=0}^{\lfloor \frac{N-2}{2} \rfloor}  \frac{d_2^n}{n!} \frac{\partial^{2n}}{\partial k^{2n}} k^{N-2}
 \\
&= (-4 d_2)^\frac{N-3}{2} k\,U\left(\frac{3-N}{2},\frac{3}{2},-\frac{k^2}{4 d_2}\right),
\label{eq:expint}
\s]
where $U$ is the confluent hypergeometric function of the second kind. 
The last equality can be shown 
by using the following relation to a hypergeometric function and comparing with its asymptotic expansion:
\[
U(a,b,z)\sim z^{-a} \, {}_2 F_0 (a,1+a-b,-z^{-1}), 
\label{eq:uasymp}
\]
where the hypergeometric function has a formal series expansion,
\[
{}_2 F_0 (a,b,x)=\sum_{n=0}^\infty \frac{(a)_n (b)_n}{n!}x^n
\]
with the Pochhammer symbol, $(a)_n=a(a+1)\cdots (a+n-1)$ ($(a)_0=1$).
For the argument in \eq{eq:expint}, the formal series stops at finite $n$, and hence \eq{eq:uasymp}
is an exact relation. One can also find 
the confluent hypergeometric function here can be expressed by an hermite polynomial.

\section{Explicit form of $S_0$}
\label{app:s0}
The $Q=\beta=0$ case was studied in \cite{Sasakura:2022axo}, and the action for this case 
is given by
\[
S_0=K_F+K_B+V_F+V_B+V_{BF},
\label{eq:s0v}
\]
where
\s[
&K_F=-\bvphit\cdot \vphit-\bpsit\cdot \vphit-\bvphit\cdot \psi_\perp+ \epsilon \bpsit \cdot \psi_\perp
-\bvphip\cdot \vphip+\bpsip\cdot \vphip+\bvphip\cdot \psip+ \epsilon \bpsip \cdot \psip,\\
&K_B=-\sigt^2-2 i \sigt\cdot\phit-\epsilon \phit^2-\sigp^2+2 i \sigp\cdot\phip-\epsilon \phip^2, \\
&V_F=-\frac{v^2}{6 \alpha} \left(
(\bpsit\cdot \vphit)^2+(\bvphit\cdot \psi_\perp)^2+2 \bpsit \cdot \bvphit \vphit\cdot \psi_\perp+2 \bpsit\cdot \psi_\perp \bvphit\cdot \vphit\right), \\
&V_B=-\frac{2 v^2}{3 \alpha} \left( 
\sigt^2 \phit^2 + (\sigt\cdot \phit)^2 \right),\\
&V_{BF}=\frac{2 i v^2}{3\alpha}
\left(
\bpsit\cdot \sigt \vphit\cdot \phit + \bvphit\cdot \sigt \psi_\perp\cdot \phit + \bpsit\cdot \phit \vphit\cdot \sigt + \bvphit \cdot \phit \psi_\perp\cdot \sigt
\right).
\label{eq:s0form}
\s]
Note that the kinetic terms of the parallel and the transverse components of the fields 
respectively have slightly different sign structures, and that the four-interactions exist only 
among the transverse components.

\section{Distribution for $N=1$ rank-one $Q$}
\label{app:neq1}
In this appendix, we derive \eq{eq:distneq1}. Ignoring all the transverse components, 
and setting $n_{\parallel}=1$ in  \eq{eq:rhodist}, \eq{eq:relwithg}, \eq{eq:swithg}, and \eq{eq:s0form}, we obtain
\[
\rho(v,q,\beta)=\pi^{-2} \alpha^\frac{1}{2} (v^4+12 \alpha \beta)^{-\frac{1}{2}} 
\exp\left[ 
\frac{-\alpha v^2 +2\alpha q v^3 -\alpha q^2 v^4}{v^4+4 \alpha \beta} \right]
(-1) \int dg d\bar \psi \cdots d\sigma\,  e^{S_{N=1}},
\label{eq:rhoneq1}
\]
where 
\s[
S_{N=1}=-g^2
-\bvphip\cdot \vphip+(a+b\, g)(\bpsip\cdot \vphip+\bvphip\cdot \psip)+ \epsilon \bpsip \cdot \psip 
-\sigp^2+2 i (a+b\, g) \sigp\cdot\phip-\epsilon \phip^2 
\s]
with $a,b$ given in \eq{eq:ab}.

The boson-fermion integration in \eq{eq:rhoneq1} produces a square root of the determinant of a two-by-two matrix. It is easy to see that the $\epsilon\rightarrow +0$ limit is smooth, and we obtain
\s[
(-1) \int dg d\bar\psi \cdots d\sigma\,  e^{S_{N=1}}&=\pi \int dg\, e^{-g^2} \left| a+b\,g \right|  \\
&=\pi \left( \sqrt{\pi}a\, {\rm Erf}\left( \frac{a}{b} \right) +b\, e^{-\frac{a^2}{b^2}} \right)
\s] 
with the error function Erf.

\section{Interactions between the $\parallel\perp_1$ and $\perp_2$ fields}
\label{app:intparaonetwo}
There are no quadratic terms containing one $\parallel\perp_1$ field and one $\perp_2$ field, because 
the index of the $\perp_2$ field cannot be contracted with $v$ or $n$. Therefore the 
$\parallel\perp_1$ fields can couple with the $\perp_2$ fields only through the four-interaction terms 
in \eq{eq:sqnbe} and \eq{eq:s0form}. 
By noting that $X_\perp\cdot Y_\perp=X_{\perp_1}Y_{\perp_1}+X_{\perp_2} \cdot Y_{\perp_2}$ for 
arbitrary fields $X,Y$, and collecting all the interaction terms between  the $\parallel\perp_1$  and 
the $\perp_2$ fields, we obtain
\s[
V_{\parallel\perp_1, \perp_2}=&\frac{8 \beta v^2}{v^4 +12 \alpha \beta} \big(
\bpsitwo \cdot \bvphitwo \psip \vphip  + \psitwo\cdot \vphitwo \bpsip \bvphip-\bpsitwo\cdot \psitwo \bvphip \vphip \\
&\hspace{2cm}
-\bpsitwo\cdot \vphitwo \bpsip \vphip-\bvphitwo \cdot \vphitwo \bpsip \psip-\bvphitwo\cdot \psitwo \bvphip \psip 
\big) \\
&+\frac{16 \beta v^2 i}{ v^4+12 \alpha \beta} \left( 
\bpsitwo \vphip+\bpsip \vphitwo+\bvphitwo \psip + \bvphip \psitwo \right) \left( \sigp \phitwo + \sigtwo \phip\right) \\
&-\frac{16 \beta v^2}{v^4+12 \alpha \beta} \left( 
\sigp^2 \phitwo\cdot \phitwo + \phip^2 \sigtwo\cdot \sigtwo + 2 \sigp\phip \phitwo\cdot \sigtwo \right) \\
&-\frac{v^2}{3 \alpha} \big( 
\bpsitwo\cdot \vphitwo \bpsione \vphione+\bvphitwo \cdot \psitwo \bvphione \psione + \bpsione\bvphione \vphitwo\cdot \psitwo \\
&\hspace{1cm}+ \bpsitwo \cdot \bvphitwo \vphione \psione+ \bpsione\psione \bvphitwo\cdot \vphitwo+\bpsitwo\cdot \psitwo \bvphione \vphione
\big) \\
&-\frac{2 v^2}{3 \alpha} \left(
\sigone^2 \phitwo^2+\sigtwo^2 \phione^2+2 \sigtwo\cdot \phitwo \sigone\phione\right) \\
&+\frac{2 v^2 i}{3 \alpha} \big(
\bpsione \sigone \vphitwo\cdot \phitwo+ \bpsitwo\cdot \sigtwo \vphione \phione+ \bvphione \sigone \psitwo\cdot \phitwo \\
&\hspace{1.5cm}+\bvphitwo\cdot\sigtwo \psione\phione+ \bpsione\phione \vphitwo\cdot\sigtwo+ \bpsitwo\cdot \phitwo \vphione\sigone \\
&\hspace{1.5cm}+\bvphione \phione \psitwo\cdot \sigtwo+ \bvphitwo\cdot \phitwo \psione\sigone \big).
\label{eq:vponetwo}
\s]

The expectation values of the $\perp_2$ fields can be taken from the large-$N$ 
Schwinger-Dyson analysis performed in \cite{Sasakura:2022iqd}.  The results 
were\footnote{To avoid duplication of notations, we use $R_{ij}$ in place of $Q_{ij}$ of \cite{Sasakura:2022iqd}.
Another thing to note is that, though we have both bosons and fermions in the current system, which is
different from the setup of \cite{Sasakura:2022iqd}, the leading-order Schwinger-Dyson analysis 
of the current system turns out to lead to the same two-fermion expectation values as \cite{Sasakura:2022iqd}. 
The reason is the presence of the supersymmetry explained below, which assures the two-boson expectation
values are just the copies of those of fermions.  }
\s[
&\langle \bpsitwo{}_a \psitwo{}_b\rangle =R_{11} \delta_{ab}, \\
&\langle \bpsitwo{}_a \vphitwo{}_b\rangle =R_{12} \delta_{ab}, \\
&\langle \bvphitwo{}_a \psitwo{}_b\rangle =R_{21} \delta_{ab}, \\
&\langle \bvphitwo{}_a \vphitwo{}_b\rangle =R_{22} \delta_{ab}, \\
&\hbox{Others}=0,
\label{eq:fermitwo}
\s] 
where, with a newly introduced parameter\footnote{As the dimension of $\perp_2$ is $N-2$, the formula 
presented in \cite{Sasakura:2022iqd} for the dimension $N-1$ of $\perp$ must be replaced with $N-2$.} $x=v^2(N-2)/(3 \alpha)$,   
\begin{itemize}
\item $0<x<1/4$
\s[
&R_{11}=\frac{-\sqrt{1-4 x}+1}{2 x\sqrt{1-4 x}},\\
&R_{12}=R_{21}=\frac{1-\sqrt{1-4 x}-4 x}{2 x \sqrt{1-4 x}},\\
& R_{22}=0, 
\label{eq:qvalsmall} 
\s]
\item $x>1/4$
\s[
&R_{11}=\frac{\sqrt{-1+4 x}}{2 \sqrt{\epsilon} x}-\frac{1}{2 x}+{\cal O}(\sqrt{\epsilon}), \\
&R_{12}=R_{21}=-\frac{1}{2 x}+\frac{\sqrt{\epsilon}}{2 x \sqrt{-1+4 x}}+{\cal O}(\epsilon^\frac{3}{2}),\\
&R_{22}=-\frac{\sqrt{\epsilon} \sqrt{-1+4 x}}{2 x}+\frac{\epsilon}{2 x} +{\cal O}(\epsilon^\frac{3}{2}).
\label{eq:qvallarge}
\s]
\end{itemize}

The two-boson expectation values can also be represented by $R_{ij}$ by assuming that a supersymmetry
is not spontaneously broken. It is easy to check that $S_0+S_{Q,\beta}$ from \eq{eq:s0v} and \eq{eq:sqbe}
are invariant under the following supersymmetry transformation:
\[
\delta \psi_a =-\phi_a,\ \delta \phi_a=\frac{1}{2} \bar \psi_a,\ \delta \varphi_a =i \sigma_a,\ \delta \sigma_a=\frac{i}{2} \bar \varphi_a,\ \delta\hbox{(others)}=0.
\]
By assuming the non-breaking of the supersymmetry, we obtain for instance a relation, $0=\langle \delta(\psi_a \sigma_b)\rangle=-\langle \phi_a \sigma_b \rangle -\langle \psi_a \frac{i}{2} \bar \varphi_b \rangle$. From such relations, we obtain
\s[
&\langle \phi_a  \phi_b\rangle =\frac{1}{2} R_{11}\delta_{ab}, \\
&\langle \phi_a \sigma_b\rangle =\frac{i}{2}R_{21} \delta_{ab}, \\
&\langle \sigma_a \phi_b \rangle =\frac{i}{2} R_{12}\delta_{ab}, \\
&\langle \sigma_a \sigma_b \rangle =-\frac{1}{2} R_{22} \delta_{ab}.
\label{eq:bosontwo}
\s]
Here, because $\phi,\sigma$ are bosons, the second and the third relations require $R_{12}=R_{21}$, 
which indeed holds in \eq{eq:qvalsmall} and \eq{eq:qvallarge}.

By putting \eq{eq:fermitwo} and \eq{eq:bosontwo} into \eq{eq:vponetwo}, we obtain
\s[
V_{\parallel \perp_1,\perp_2}(R)=&-A_1 \left( R_{22} \bpsip \psip+R_{12}(\bpsip \vphip +\bvphip \psip)+R_{11}\bvphip \vphip\right)\\
&+A_1 \left( R_{22} \phip^2-2 i R_{12} \phip\sigp-R_{11}\sigp^2 \right)\\
&-A_2 \left( R_{22} \bpsione \psione+R_{12}(\bpsione \vphione +\bvphione \psione)+R_{11}\bvphione \vphione\right)\\
&+A_2 \left( R_{22} \phione^2-2 i R_{12} \phione\sigone-R_{11}\sigone^2 \right),
\label{eq:vp12q}
\s]
where $A_i$ are given in \eq{eq:Ai}.
Adding \eq{eq:vp12q}  to \eq{eq:kineticp1}, we obtain the full kinetic terms of the $\parallel\perp_1$ fields.
In particular the fermionic part has the form, 
\[
K_{\parallel \perp_1}+V_{\parallel \perp_1,\perp_2}(R) = \bar  \psi_{\parallel\perp_1} H \psi_{\parallel\perp_1}
+\hbox{bosonic part},
\]
where $\bar \psi_{\parallel\perp_1}=(\bpsip, \bvphip, \bpsione,\bvphione),
\psi_{\parallel\perp_1}=(\psip, \vphip, \psione,\vphione)$ and $H$ is given in \eq{eq:defofH}. 
Because of the supersymmetry, the bosonic
part has essentially a parallel structure as $H$.

\section{Exact expression of $Z_{\perp_2}$}
\label{app:exactz1}
$Z_{\perp_2}$ can be taken from \cite{Sasakura:2022axo}, because it is the same as 
the partition function of the transverse components of the fields in  \cite{Sasakura:2022axo},
which is denoted as $G_N$ there. A point to note is that, while the dimension of the transverse directions there 
is $N-1$, the dimension of $\perp_2$ of this paper is $N-2$. Therefore we have to deduct $N$ by 
one\footnote{See also the footnotes associated to \eq{eq:zapp2} and \eq{eq:fermitwo}.}, when we take a result 
from \cite{Sasakura:2022axo}.  
In our current case of $N=9$, this corresponds to $G_{N=8}$, and therefore
\s[
Z_{\perp_2}^{N=9}&= G_{N=8} \\
&=\pi^{\frac{13}{2}} \Bigg( \frac{
   \sqrt{2}
     e^{-\frac{1}{8z}} (1 + 210 z^2 - 2100 z^3 + 12600 z^4 + 25200 z^5)}{
   15 z^\frac{3}{2}} 
   \\
   &
   \hspace{5cm}
   + (1 - 42 z + 420 z^2 - 840 z^3) \,\gamma\left[\frac{1}{2},\frac{1}{8 z}\right]
    \Bigg),
    \label{eq:hforeven}
\s]
where $z=v^2/(6 \alpha)$, and $\gamma\left[1/2,y\right]$ is 
the lower incomplete gamma function with index $1/2$, which is related to the error function by
\[
\gamma\left[\frac{1}{2},y\right]=\sqrt{\pi}\, \hbox{Erf}\left(\sqrt{y}\right).
\]

\section{Proposition 2 of \cite{AMN}}
\label{app:amn}
In this appendix we compare our threshold value $\tal$ with the value given in Proposition 2 of \cite{AMN}.

In \cite{AMN} with their notations, the random tensor $Y$ with a background is given by 
\s[
Y=\lambda u^{\otimes k} +\frac{1}{\sqrt{2 N}} W,
\s]
where $|u|=1$, $W=\sum_\pi G^\pi/k!$ with $G_{i_1\cdots i_k}\sim N(0,1)$, and $\pi$ denotes permutations. 
By taking $k=3$ and computing the standard deviations of $W_{abc}$, we find 
\[
W_{abc}\sim N\left(0,1/\sqrt{d_{abc}}\right),
\]
where $d_{abc}$ is the degeneracy factor defined in \eq{eq:degen}. 
In our case, $C_{abc}\sim N(0,1/\sqrt{2 \alpha d_{abc}})$, and therefore 
\[
\alpha=N
\]
is taken in \cite{AMN} in our notation. $q=\lambda$ is also taken.
Therefore,
\[
\tilde \alpha=\frac{\alpha q^2}{N}=\lambda^2.
\]
Proposition 2 states
\[
\lambda_c^2=\frac{(k-1)^{k-1}}{2 k (k-2)^{k-2}}=\frac{2}{3}
\]
for $k=3$.
Therefore the threshold value of \cite{AMN} corresponds to $\tal_c=2/3$.

\section{Maximums of $S_\infty^\gamma$}
\label{app:maxs}
In this section we will prove that $S_{\infty}^\gamma<0$ for $0<\tv<\infty$ and $-1\leq \np<1$.
Firstly, 
\[
(v^4+4 \tal \tbe) ( v^4+ 12 \tal \tbe) S_\infty^\gamma =
-\tal \tv^2 (\tv^4 - 2 \np^3 \tv^5 + (3 \np^4 - 2 \np^6) \tv^6 + 
   12 \tal \tbe (1 - 2 \np^3 \tv + \np^4 \tv^2)).
\]
Therefore the statement is equivalent to prove the positivity of the quantity in the parentheses:
\s[
&\tv^4 - 2 \np^3 \tv^5 + (3 \np^4 - 2 \np^6) \tv^6 + 12 \tal \tbe (1 - 2 \np^3 \tv + \np^4 \tv^2) \\
&\ =\left(1 - 2 \np^3 \tv+\np^4 \tv^2 \right)\tv^4  + 2 \np^4(1 -  \np^2) \tv^6 + 12 \tal \tbe (1 - 2 \np^3 \tv + \np^4 \tv^2).
\s]
For $\np<1$, we find 
\[
1 - 2 \np^3 \tv+\np^4 \tv^2 > 1-2 \np^2 \tv +\np^4 \tv^2 =(1-\np^2 \tv)^2\geq 0. 
\] 
Similarly one can prove that the quantities in the other parentheses are larger than zero.

\end{document}